\documentclass[preprint2]{emulateapj}
\slugcomment{To appear in Astrophysical Journal}
\shorttitle{Hubble Sequence at $z \sim 2$}
\shortauthors{Lee et al.}

\begin{document}

\title{CANDELS: The correlation between galaxy morphology \\
  and star formation activity at $z\sim 2$}

\author{Bomee Lee\altaffilmark{1}, Mauro Giavalisco\altaffilmark{1}, Christina C. Williams\altaffilmark{1}, Yicheng Guo\altaffilmark{2}, \\
  Jennifer Lotz\altaffilmark{3}, Arjen Van der Wel\altaffilmark{4}, \\ 
  Henry C. Ferguson\altaffilmark{3}, S.M. Faber\altaffilmark{2}, Anton Koekemoer\altaffilmark{3}, Norman Grogin\altaffilmark{3}, Dale Kocevski\altaffilmark{8}, \\
  Christopher J. Conselice\altaffilmark{5}, Stijn Wuyts\altaffilmark{6}, Avishai Dekel\altaffilmark{7}, Jeyhan Kartaltepe\altaffilmark{9}, Eric F. Bell\altaffilmark{10}}

\email{bomee@astro.umass.edu}

\altaffiltext{1}{Department of Astronomy, University of Massachusetts, Amherst, MA 01003, USA}
\altaffiltext{2}{University of California Observatories/Lick Observatory, University of California, Santa Cruz, CA 95064, USA}
\altaffiltext{3}{Space Telescope Science Institute, 3700 San Martin Drive, Baltimore, MD 21218, USA}
\altaffiltext{4}{Max-Planck Institut $f\ddot{u}r$ Astronomie, $K\ddot{o}$nigstuhl 17, D-69117, Heidelberg, Germany} 
\altaffiltext{5}{University of Nottingham, School of Physics and Astronomy, Nottingham, NG7 2Rd UK}
\altaffiltext{6}{Max-Planck Institut $f\ddot{u}r$ Extraterrestrische Physik, Giessenbachstrasse, D-85748 Garching, Germany}
\altaffiltext{7}{Racah Institute of Physics, The Hebrew University, Jerusalem 91904, Israel}
\altaffiltext{8}{Department of Physics and Astronomy, University of Kentucky, Lexington, KY40506, USA}
\altaffiltext{9}{NOAO-Tuscon, 950 North Cherry Ave., Tucson, AZ 85719, USA}
\altaffiltext{10}{Department of Astronomy, University of Michigan, 500 Church St., Ann Arbor, MI 48109, USA}

\begin{abstract}
  We discuss the state of the assembly of the Hubble Sequence in the mix of
  bright galaxies at redshift $1.4< z\le 2.5$ with a large sample of 1,671
  galaxies down to $H_{AB}\sim26$, selected from the HST/ACS and WFC3 images of
  the GOODS--South field obtained as part of the GOODS and CANDELS
  observations. We investigate the relationship between the star formation
  properties and morphology using various parametric diagnostics, such as the
  $S\acute{e}rsic$ light profile, Gini ($G$), $M_{20}$, Concentration ($C$),
  Asymmetry ($A$) and multiplicity ($\Psi$) parameters. Our sample clearly
  separates into massive, red and passive galaxies versus less massive, blue and
  star forming ones, and this dichotomy correlates very well with the galaxies'
  morphological properties. Star--forming galaxies show a broad variety of
  morphological features, including clumpy structures and bulges mixed with
  faint low surface brightness features, generally characterized by disky-type
  light profiles.  Passively evolving galaxies, on the other hand, very often
  have compact light distribution and morphology typical of today's spheroidal
  systems. We also find that artificially redshifted local galaxies have a
  similar distribution with $z\sim2$ galaxies in a $G$--$M_{20}$ plane. Visual
  inspection between the rest-frame optical and UV images show that there is a
  generally weak morphological k-correction for galaxies at $z\sim2$, but the
  comparison with non-parametric measures show that galaxies in the rest-frame
  UV are somewhat clumpier than rest-frame optical. Similar general trends are
  observed in the local universe among massive galaxies, suggesting that the
  backbone of the Hubble sequence was already in place at $z\sim 2$.
\end{abstract}


\section{Introduction}

Galaxy morphology in the local universe provides significant information about
the formation and evolution of galaxies. Massive galaxies in the nearby
universe are well described by the Hubble sequence, which correlates with the
dominance of galaxy's central bulge, surface brightness and colors.  Hubble
types are also broadly correlated with physical parameters, such as the star
formation rate and dynamical properties \citep{rob94, bla03}. In the classical
picture, late-type spiral galaxies are active star--forming structures with
flattened, gas rich, rotationally supported exponential disks, while
early-type galaxies are more luminous, massive and quiescent systems,
supported by stellar velocity dispersion and have smooth elliptical isophotes
with a so--called ``de Vaucouleurs" (or similar) light profile.  In a
color-magnitude (or mass) diagram of the local universe, the early- and
late-type galaxies occupy two distinct regions, known as the red sequence and
blue cloud, respectively (Blanton et al. 2003, Baldry et al. 2004).  The red
sequence consists of mostly non-star-forming galaxies with a bulge dominated
structure and colors indicative of an old stellar population.  In contrast,
the galaxies lying in the blue cloud have different star formation properties,
including blue star-forming stellar populations and mostly disk-like
structures.

The key question of how and over what time--scale the Hubble Sequence has
formed remains unanswered. At a basic level, exploring the origin of the
Hubble sequence can simply be done by investigating if high-redshift galaxies
have distributions of morphological types (early- and late-type) and
star-forming properties that resemble those in the nearby universe. Several
surveys that have used the Hubble Space Telescope (HST) have observed that the
properties of galaxies at $z\sim1$ are broadly consistent with those in the
local universe (Bell et al. 2004: GEMS, Papovich et al. 2005: HDFN, Cassata et
al. 2007 : COSMOS).  However, the morphological analysis is still
controversial at the peak epoch of star-formation activity ($z\sim2-3$). Until
relatively recently, most studies of galaxy morphologies at $z>2$ have been
performed at rest--frame ultraviolet (UV) wavelengths using optical imagers
(such as HST/WFPC2 and HST/ACS). These works found that irregular or peculiar
structures appear more common, and traditional Hubble types do not appear to
be present at these epochs \citep{gia96a,gia96b, ste96, low97, lot04, pap05,
  lot06, rav06, law07, con08}.  This is generally explained as due to the fact
that UV radiation predominantly traces emission from the star-forming regions
\citep{dic00}, which tend to be more clumped and irregularly distributed than
older stellar populations, and also by the fact that quenched galaxies were
missing from the optical images. The rest-frame optical regime is a better
probe of the overall stellar distribution in galaxies, and early
near--infrared (NIR) observations with {\it HST} and NICMOS of star--forming
galaxies at $z>2$ from UV selected samples found that their morphology remains
generally compact and disturbed also at rest-frame optical wavelengths and
bear no obvious morphological similarities to lower redshift galaxies
\citep{pap05, con08}. Interestingly, however, \cite{kri09} showed that 19
spectroscopically confirmed massive galaxies ($>10^{10.5}M_{\odot}$) at $z\sim
2.3$ are clearly separated into two classes as a blue cloud with large
star-forming galaxies, and a red sequence with compact quiescent galaxies.
Unlike late--type galaxies, early-type galaxies (ETGs) have been used to
investigate the cosmic history of massive galaxies in many studies
(e.g. Renzini (2006), and references therein) due to their simple elliptical
morphologies and passively evolving stellar populations. At $z<1$, there is
general consensus that the majority of massive ETGs ($M > 10^{11}M_{\odot}$)
were already in place at $z\sim 0.8$, with a number density comparable to that
of local galaxies \citep{cim02, im02}.  A number of studies have reported the
emergence of massive and compact galaxies by $z\sim 2-3$, which are already
quenched ETGs \citep{cim04, dad05, tru06, tru07, van08, cas08, val10, cas10}.
The number density of these galaxies rapidly increases, by a factor of five,
from $z\sim2$ to $z\sim1$, and they are up to 5 times more compact in size
than local ones with similar mass \citep{cas11, cas13}. Recent works have
suggested, however, that a large fraction of, and possibly even all, massive,
quiscent galaxies at $z\sim2$ are disk dominated \citep{vdw11}. While the
observation of such disks at $z>2$ is based on morphological analysis alone,
typically distribution of apparent elongation, with no dynamical measures at
present, the existence a sizeable fraction of compact disks at $z>2$ among
massive, passive galaxies \citep{van11,wan12,bru12} suggests that the
physical mechanism responsible for quenching star-formation may be distinct
from the process responsible for morphological transformation.

Recently, studies of the morphologies of $z\sim2$ galaxies have advanced 
using the high resolution NIR Wide-Field Camera 3 (WFC3). \cite{szo11} 
found a variety of galaxy morphologies, ranging from large, blue, disk-like 
galaxies to compact, red, early-type galaxies at $z\sim2$ with 16 massive 
galaxies in the Hubble Ultra Deep Field (HUDF). \cite{cam11} also studied the 
rest-frame UV and optical morphologies with $1.5 <z<3.5$ galaxies 
determined by the YHVz color-color selection in the HUDF and the 
Early Release Science (ERS) field, and confirmed previous studies by 
showing in particular the presence of regular disk galaxies, which 
have been missing in previous studies, either because they are not
detected at the available sensitivity or because their appearance is irregular
at rest-frame UV wavelengths. The results from these studies are generally
interpreted as possible evidence at $z\sim 2$, at least among the brightest
galaxies at that epoch, of the general correlations between spectral types and
morphology that today define the Hubble Sequence.  The most important
limitations of these works are that they are based on very small samples,
which are not statistically significant (less than 20 galaxies) and not
homogeneously selected, and their morphological analysis was restricted to only
$S\acute{e}rsic$ profile fitting or visual classifications.

Significant improvements are now possible using larger samples of panchromatic
images from the CANDELS (Cosmic Assembly Near-infrared Extragalactic Legacy
Survey) observations. \cite{wuy11} investigated how the structure of galaxies
($S\acute{e}rsic$ index and size) depends on galaxy position in the
SFR--stellar mass diagrams since $z\sim2.5$, specifically showing strong
trends of specific star formation rate (SSFR) with $S\acute{e}rsic$ index 
using large data sets combined in 4 different fields, 
COSMOS, UDS, GOODS-South and North. Although the
$S\acute{e}rsic$ index, measured by fitting a single $S\acute{e}rsic$ profile
to a galaxy, is the most common approach to analyzing galaxy morphology, it is
also useful to study morphologies with non-parametric measures such as Gini
($G$) \citep{abr03}, $M_{20}$ (Lotz et al. 2004), multiplicity 
($\Psi$) (Law et al. 2007) and CAS (Abraham et al. 1996, Conselice et al. 2003) 
since not all galaxies are described by smooth and symmetric profiles. 
In a recent CANDELS paper, \cite{wan12} used Gini ($G$), $M_{20}$ and visual 
classifications to identify a correlation between morphologies and star-formation 
status at $z\sim2$, showing two distinct populations, bulge-dominated quiescent 
galaxies, and disky or irregular star-forming galaxies, though they only use massive
galaxies ($M>10^{11}M_{\odot}$). Recent panchromatic surveys such as CANDELS
hold the promise of significant progress in investigating galaxy structures at
high redshift because they combine sensitive {\it HST} morphology at
rest-frame UV and optical wavelengths with the depth and accuracy of
space--borne photometry. The CANDELS project also adds coverage of a
substantial amount of sky, which results in samples whose size and dynamic
range in mass are about one order of magnitude , or more, larger than in
previous works. In this study, we extend previous results using a
statistically significant sample (1,671 galaxies) down to a lower mass limit
($M > 10^{9}$ M$_{\odot}$ at $z\sim 1.4$ and $10^{10}$ M$_{\odot}$ at $z\sim
2.5$, specifically for passive galaxies) and using various morphological
parameters (non-parametric diagnostics such as $G$, $M_{20}$, $\Psi$,
Concentration (C) and Asymmetry (A), and the parametric $S\acute{e}rsic$ light
profiles, characterized by the $S\acute{e}rsic$ index (n) and half-light radius $R_{e}$), 
as well as visual inspection.

The combination of high--angular resolution and sensitivity afforded by the
{\it HST}/ACS and WFC3 images with the relatively large size of the sample
allows us to probe in a statistical fashion the correlations between galaxy
structures and star-formation activity at $z\sim2$, i.e. the epoch when the
cosmic star--formation activity reached its peak, to their counterparts in the
local universe. The structure of this paper is as follows. The optical and
infrared data and selection of our galaxy sample are introduced in Section
2. The rest-frame color and mass distributions are described in Section 3. We
present the analysis of galaxy morphologies in the rest-frame optical using
the distribution of non-parametric approaches, as well as $S\acute{e}rsic$
index and half-light radius in Section 4. Comparison with galaxies from the local
universe is shown in Section 5 and the results of a comparison of rest-frame
UV morphologies with rest-frame optical are presented in Section 6. Finally,
we conclude with a discussion of our results and compare them to other studies
in Section 7 and summarize our results in Section 8.

\section{Data and Sample Selection}

All the data used in this work come from the observations acquired during the
GOODS (Giavalisco et al.\ 2004) and CANDELS (Grogin et al.\ 2012; Koekemoer et
al.\ 2012) projects in the GOODS--South field, including both space--born
(Chandra, Hubble and Spitzer) as well as ground--based (VLT) data. The CANDELS
{\it HST} observations, including the details of the data acquisition,
reduction and calibration, source identification and photometry extraction,
are thoroughly described elsewhere (see Grogin et al.\ 2012; Koekemoer et
al.\ 2012; Guo et al.\ 2013); here we briefly review key features of the WFC3
images that are relevant to this work. The {\it HST} component of CANDELS
consists of a Multi-Cycle Treasury program to image five distinct fields
(GOODS-North and -South, EGS, UDS and COSMOS) using both WFC3 and ACS.  The
whole project is organized as a two--tier Deep+Wide survey. The CANDELS/Deep
survey covers about 125 square arc minutes to $\sim 10$-orbit depth within
GOODS-North and -South \citep{gia04} at F105W(Y), F125W(J) and F160W(H), while
the Wide survey covers a total of $\sim 800$ square arc minutes to $\sim
2$-orbit depth within all five CANDELS fields.  In this study, we use the
4-epoch ( about 2 orbits) CANDELS F160W(H-band) imaging that covers about 115
square arc minutes ($\sim 2/3$ of the whole GOODS-S) including the GOODS-S
deep region, plus the ERS \citep{win11}.  This survey has a $5\sigma$ limiting
depth of $H_{AB} = 27.7$, and a drizzled pixel scale of $0.06"$. A number of
photometric catalogs exist based on CANDELS data, and here we use one where
sources have been detected using the package SExtractor \citep{ber96} in the
WFC3 H--band images, and multi--wavelength photometry has been obtained using
a software package with an object template-fitting method (TFIT, Laidler et
al. 2007). This catalog includes photometry from the {\it HST}/ ACS and WFC3
images in the BVizYJH bands; from VLT/VIMOS U and VLT/ISAAC Ks images; and
from the Spitzer/IRAC images at 3.6, 4.5, 5.8 and 8.0 $\mu$m (Guo et
al.\ 2013).

We identify galaxies at $1.4< z\le 2.5$ with a broad range of star--formation
properties, from passive to star forming, and with different levels of dust
obscuration using photometric redshifts and SSFR estimated by fitting the
CANDELS broad-band rest--frame UV/optical/NIR spectral energy distribution
(SED) to spectral population synthesis models (hereafter, SED
sample). Additionally, for comparison, we also select samples of galaxies
using the BzK technique, a color selection based on the (B-z) vs. (z-K)
color-color diagram, widely used to identify galaxies at $z\sim2$ relatively
independently of their star--formation activity and dust obscuration
properties \citep{dad04,dad07}. While characterized by some contamination by
AGNs and low--redshift interlopers, as well as incompleteness to very young
and dust--free star--forming galaxies (see \cite{dad04}), the BzK selection
is overall quite effective and particularly economic in that it only requires
the acquisition of three photometric bands. In contrast, the SED selection is
observationally much more expensive because it requires a large number of
photometric bands to yield robust photometric redshifts as well as robust
measures of the stellar population properties, i.e.  stellar mass,
star--formation rate, dust obscuration and age of the dominant stellar
population. For the same reason, however, it is less prone to the effects of
photometric scatter and characterized by a higher degree of completeness than
the BzK criterion.

In view of the fact that in CANDELS the two GOODS fields have deeper and fully
panchromatic images relative to the other fields of the survey, here we use
the SED sample as our primary data set for our study, and compare it with the
BzK sample to test if they yield similar conclusions about the general
morphological properties of the galaxies mix at $z\sim 2$. Such a comparison,
which at this level of sensitivity can only be made in the GOODS fields, is
particulary useful for those other fields where data for selecting galaxies by
means of SED fitting are not available or do not have sufficient wavelength
coverage and/or sensitivity for accurate results.  In our particular case,
since the BzK sample is bsed on the ground--based K-band images, which are
significantly shallower ($5\sigma$ limiting magnitudes of Ks=24.4) than the
WFC3 images, the depth, and hence the size, of the sample is smaller than the
SED one. However, since the efficiency and simplicity of the BzK selection
criteria offer a distinct advantage in other fields of the CANDELS survey,
where the rich complement of photometry of the GOODS--South field is not
available, the knowledge of the relative performance and possible limitations
of both selection criteria will be very useful.

\begin{figure*}
\begin{center}
\epsscale{0.9}
\plotone{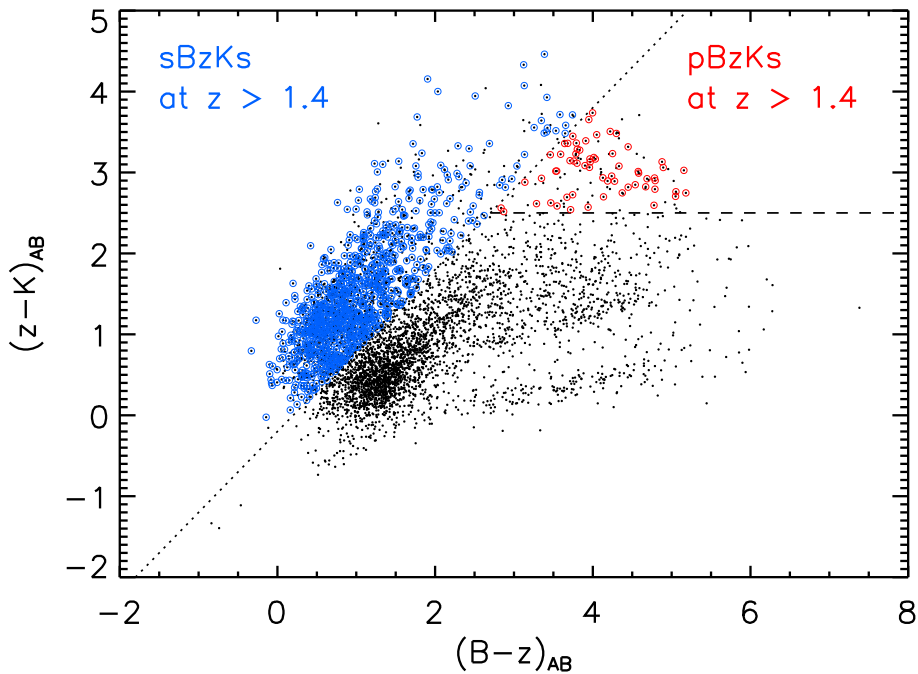}
\caption{\small (z-K) vs. (B-z) diagram for the z-band selected sources in the GOODS-
South field of the HST/ACS [black points] with $S/N_{k} > 7$ and $S/N_{z} > 10$. 
Sources above the dotted line are classified as the star-forming galaxies [sBzKs] 
and sources between the dotted and dashed lines are the passively evolving galaxies [pBzKs]. 
1043 BzK galaxies are detected in WFC3/F160w (H-band) observation, using epoch 4 
of the CANDELS. The blue and red circles identify these 981 sBzKs and 62 pBzKs, 
respectively [BzK sample]. }
\label{fig:bzk}
\end{center}
\end{figure*}

\subsection{The Galaxy Mix at $z\sim 2$: Photometric Redshift and SED--fitting selection} 

Measures of the stellar mass, star--formation rate, dust obscuration and age
of the dominant stellar population have been obtained by Guo et al.\ (2013)
using the TFIT panchromatic catalog of the GOODS--South field (see also Guo et
al.\ 2011, 2012, where key results and features of the SED fitting have been
discussed). Prior to carrying out the SED fitting, photometric redshifts have
been measured for all galaxies from the 13--band
UBVizYJHKsI$_1$I$_2$I$_3$I$_4$ photometry using the PEGASE 2.0 spectral
library templates \citep{fio97}, as well as the available sample of 152
spectroscopic redshifts (about 4\% of our final sample) as a training set. In
the redshift range considered here the CANDELS photometric redshifts are of
good quality, as verified by comparing them against available spectroscopic
ones. This comparison yields a mean and scatter ( $1\sigma$ deviation after
$3\sigma$ clipping) in our photometric redshift measurement of 0.0005 and
0.03, respectively.

The properties of the dominant stellar populations are subsequently derived by
fitting the observed SED to the spectral population synthesis models by fixing
the redshift to the photometrically derived value and using the redshift
probability function, $P(z)$, to calculate the errors from a Monte Carlo
bootstrap.  The multi--band photometry is fit to the updated version of the
\cite{bru03} spectral population synthesis library with a Salpeter initial
mass function. We use either a constant star formation history or an
exponentially declining one ($e^{-t/\tau}$), depending on which functions
result in a smaller $\chi^{2}$ with the data. The Calzetti law \citep{cal00}
is used for the dust obscuration model together with the \cite{mad95}
prescription for the opacity of Inter galactic medium (IGM) (see Guo et
al.\ 2013 for a full description of the procedure). In the redshift range $1.4
< z\le 2.5$, arbitrarily (but inconsequentially) chosen to reproduce that of
the BzK selection criteria (see \cite{dad07}), the photo--z plus SED fitting
procedures yield 3,542 galaxies with signal to noise ratio in the H-band
$(S/N)_{H}>10$ (hereafter, SED sample).

Star--forming and passive galaxies are defined based on the measure of the
SSFR, namely the ratio of the star--formation rate to the stellar
mass. Specifically, we define passive galaxies as those with
$SSFR<0.01~Gyr^{-1}$, and using this criterion we find 105 passive galaxies
and 3,437 star--forming ones out of the 3,539 comprising the SED sample. Thus,
with our cut on the SSFR, 3\% of all galaxies at $z\sim 2$ are classified as
passive.

\subsection{The Galaxy Mix at $z\sim 2$: The BzK Selection}

We have constructed the BzK sample by adopting the BzK color--color selection
by \cite{dad04}, where galaxies of various ``spectral types" are identified by
their position in the $(B-z)$ versus $(z-K)$ color-color diagram. The BzK
selection is widely used to investigate the evolution of galaxies at $z\sim2$
\citep{dad05, red05, dad07,lin12, fan12, yum12}.  To the extent that the average obscuration
properties of the star--forming galaxies are well described by the Calzetti
(2000) obscuration law, this rest UV/Optical color selection is sensitive to
galaxies at $1.4<z\le2.5$ with a significantly broader range of dust
obscuration than the UV selection alone (e.g. Reddy et al. 2009, 2010; also
Guo et al. 2011 for a discussion). It is also sensitive to passively evolving
galaxies in a similar redshift range, which the UV selection misses
altogether. As in any selection of distant galaxies that is based on colors,
however, the details of the redshift distribution of the selected galaxies
depend on the relative sensitivity of the images and the shape of the adopted
bandpasses.
\begin{figure}
\begin{center}
\epsscale{1.0}
\plotone{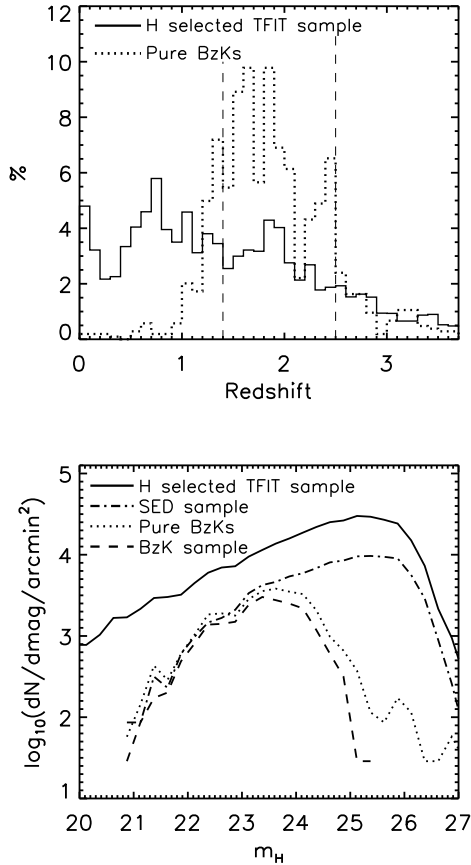}
\caption{\small Comparison of the photometric redshift distribution 
of the BzK sample to that of the SED sample (Top) and the number counts of 
SED sample, of the two BzK samples, and
also of the whole H--band selected TFIT sample for comparison (Bottom). 
The thick solid line represents the H-band selected TFIT sample (e.g. SED 
sample without redshift limitation), and dotted line represents the ``Pure BzK". 
The redshift window, $ 1.4 < z \le 2.5$, is described as vertical dashed lines in 
the top panel. In the Bottom panel, the dot and dash dot lines are for the BzK 
sample (BzKs at $z\sim2$) and SED sample, respectively. }
\label{fig:numcount}
\end{center}
\end{figure}
The original BzK criteria were implemented using a sample where source
detections were carried out in the K-band images, since these had sufficient
sensitivity and were such that every galaxy detected at least in the z-band
was detected in the K one with higher S/N. Since this is not the case with our
data, where the {\it HST}/ACS z-band image is much deeper than the
ground--based VLT/ISAAC Ks band image even for the reddest SED considered
here, we contruct our BzK samples from the ACS z-band selected source catalog
\citep{gia04}, where we further require $(S/N)_{K} > 7.0$ in the K-band and
$(S/N)_{z}>10.0$ in the z-band to ensure robust color measurements. We then
use the selection criteria introduced by \cite{dad04} as shown in
Figure~\ref{fig:bzk}: 
\begin{eqnarray}
BzK \equiv (z-K)-(B-z) \ge -0.2 \qquad \nonumber \\
\textrm{for star-forming galaxies (sBzKs) and } \nonumber \\
BzK < -0.2 ~~\cap ~~(z-K) > 2.5 \qquad  \nonumber \\
\textrm{for passively evolving galaxies (pBzKs)}\nonumber
\end{eqnarray}

Out of a total of 1,043 BzK galaxies, we find 981 sBzKs (blue circles in
Figure~\ref{fig:bzk}) and 62 pBzKs (red circles), namely 6\% of the sample is
made of passive galaxies. This fraction is twice as large as the one of the
SED sample, and the reason is that the BzK selection defines galaxies as
passive solely based on their colors, while in the SED sample galaxies are
defined as passive based on the SSFR. If the threshold were defined as
SSR$<0.16$ Gyr$^{-1}$, then the SED sample and BzK would both have the same
6\% fraction of passive galaxies. Finally, it is important to keep in mind
that all BzK galaxies also have detection in the WFC3/F160w CANDELS images,
which we have used to analyze their rest--frame optical morphology.

Compared to the SED selection, the BzK selection is relatively easy and
economical to implement, requiring only three photometric bands, and it is
largely independent of dust reddening. The combination of photometric scatter
(which depends on the sensitivity of the data) and the intrinsic variance of
galaxies' SED, however, result in some degree of contamination by interlopers
(i.e. galaxies that are not in the targeted redshift range) as well as of
incompleteness, namely loss of galaxies that are scattered away from the
selection windows. For the same reasons, the separation between passive
galaxies and star--forming ones suffers from scatter, in the sense that
dust--reddened star--forming galaxies might be observed in the selection
window of passive ones and vice versa (see, e.g. Daddi et al. 2004, 2005,
2007).

\begin{figure*}
\begin{center}
\epsscale{1.0}
\plotone{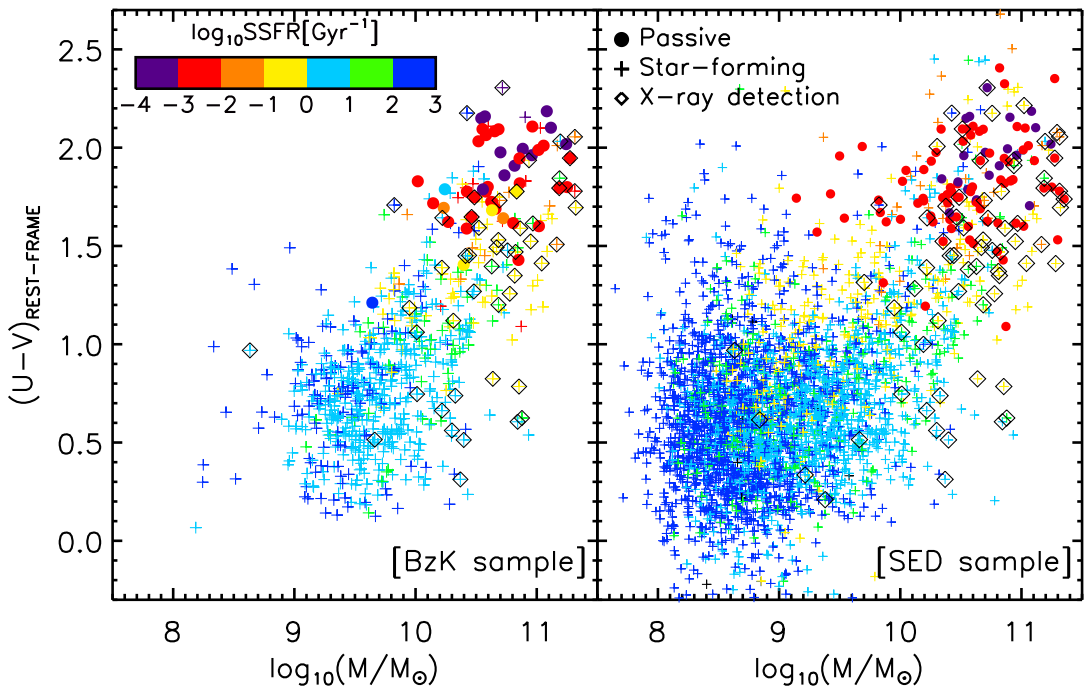}
\caption{\small The rest-frame U-V color versus Stellar mass for galaxies of both BzK 
(left) and SED (right) samples at $z\sim2$. The crosses and circles represent the 
star-forming (sBzKs) and passive (pBzKs), respectively. The rest-frame colors 
of the passive (pBzK) galaxies span a much smaller range than star-forming (sBzK) 
ones, with the two samples having distinct color distributions. The color-coding reflects 
the specific star formation rate (SSFR) defined as star formation rate divided by 
stellar mass (right panel). Most massive pBzKs have $SSFR < 0.01~ Gyr^{-1} $ 
which means they are rarely star-forming and intrinsically red in contrast to blue 
sBzKs with higher SSFR and lower mass. Diamond symbols 
represent the 70 x-ray detected galaxies (50 for BzK sample).}
\label{fig:CM}
\end{center}
\end{figure*}

To diminish the contamination by interlopers from outside the redshift range
considered here and how it affects our morphological analysis, we can use the
photometric redshift to restrict our sample to galaxies with $1.4< z\le
2.5$. This further cut serves two purposes. First, it filters away the high--z
tail of our BzK sample, which very likely results from the combination of the
relatively shallow depth of the ACS B--band images, and the fact that the
sample is z--band selected. The cut also serves the purpose of creating our
reference cosmic epoch to assess the state of galaxy morphology evolution.
This leaves a final BzK sample of 736 galaxies down to $H\sim25$, of which 46
are classified as passively evolving (pBzK), i.e. 6.3\% of the sample, and 690
are star--forming (sBzK) galaxies. We explicitly note that using the
photometric redshifts to eliminate likely interlopers has left the passive
fraction essenctially unchanged. In the following we will refer to this
photo--z filtered sample simply as the ``BzK sample'', while the original
sample will be called the ``pure BzK'' one.  The top panel of the
Figure~\ref{fig:numcount} compares the photometric redshift distribution of
the BzK sample to that of the whole H--band selected TFIT sample (e.g. SED
sample not restricted by redshift), while the bottom one shows the number
counts of the SED sample, of the two BzK samples, and also of the whole
H--band selected TFIT sample for comparison. It can be seen from the
photometric redshift histograms that the redshift distribution of BzKs is
tapered at both ends of the corresponding selection window as a result of the
color cuts built in the selection criteria. Clearly, this is not presented in
the SED sample. It can also be seen that the number counts of the SED and BzK
samples are very similar in shape, especially at $H < 25$, the former being
slightly larger than the latter simply due to the larger completeness and the
fact that the redshift distribution is not set by color cuts.  Since the
magnitude (mass) distribution is not similar especially at the faint end, we
cut BzK and SED samples with $M>10^{9}M_{\odot}$ ( $H\lesssim 26$ : over the
90\% completeness limit of the CANDELS $H$ band in the GOODS-S Deep field) to
study and compare the morphologies directly. As we shall see later, the
morphological properties of the SED and BzK samples are essentially identical,
suggesting they are both representative of the mix of bright galaxies at
$z\sim 2$.

\section{ Color-mass diagram at $z\sim2$}

\begin{figure}
\begin{center}
\epsscale{1.0}
\plotone{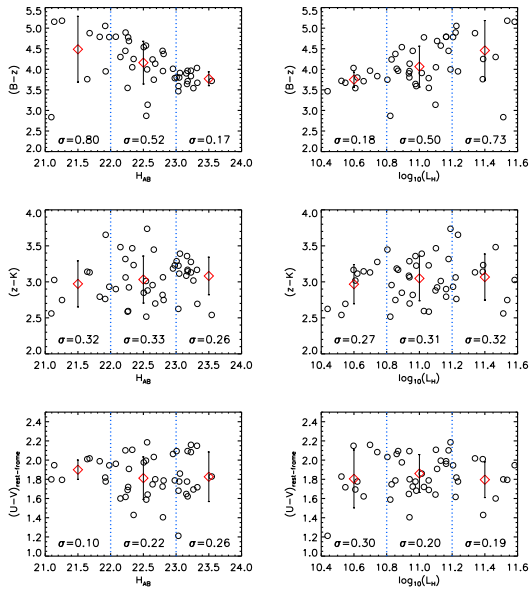}
\caption{\small The observed colors (B-z and z-K) and rest-frame color (U-V) versus 
apparent ($H_{AB}$) and absolute magnitudes ($log(L_{H})$) for pBzK galaxies. 
The $\sigma$ values and error bars represent the standard deviation of colors.}
\label{fig:p_sigma}
\end{center}
\end{figure}

In Figure~\ref{fig:CM}, we report the distribution of rest-frame U--V (3730\AA
-- 6030\AA) color versus stellar mass for the BzK (left) and SED (right)
samples. The blue crosses and red circles represent the star-forming (sBzK)
and passive (pBzK) galaxies, respectively. The rest-frame colors are measured
using the best-fitting SED template, which is scaled to pass through the closest
observed photometric point for each rest-frame wavelength we consider to
derive the k-correction and subsequently the rest-frame
magnitude. Figure~\ref{fig:CM} shows that there is a distinct difference in
the color-mass diagram between star-forming and passive galaxies. In both our
samples, the colors of passive galaxies (pBzKs) span a much smaller range than
those of star-forming ones (sBzKs). An important question to answer,
therefore, is whether or not the limited excursion of the intrinsic colors of
the pBzKs is simply the result of their selection and not due to the
characteristics of their SED. After all, these galaxies are selected
specifically to be red, namely to have the observed colors expected from
passively evolving galaxies (or galaxies with a relatively small specific
star formation rate) observed at $1.4 < z \le 2.5$. To test this possibility,
we have compared the scatter of the observed colors and of the intrinsic
colors of our pBzK sample in bins of both apparent and absolute magnitude,
which is shown in Figure~\ref{fig:p_sigma}. As it can be seen, the scatter
{\it always} increases when going from the intrinsic colors to the observed
ones, as one would expect in a sample with a relatively large dispersion of
redshift. Moreover, the pBzKs occupy a significantly smaller range of stellar
mass, and, at the same time, the two types occupy a disjoint range of SSFR
(color-coding of points at Figure~\ref{fig:CM}). Taken all
together, this is evidence that while pBzKs are selected to be red, thus
covering a restricted range of both the {\it observed} B-z and z-K colors,
their {\it intrinsic} colors are all very similar, since they span a range
significantly smaller than the observed ones, denoting a similarity of
physical properties. This conclusion is further reinforced by their small
range of mass, since the color selection does not in principle set any
restrictions on the stellar mass.  

We note that although our star-forming and passive samples seperate well in
Figure~\ref{fig:CM}. a small fraction of massive star-forming galaxies overlap
with the passive ones. The majority of those massive star-forming galaxies are
red due to UV colors which are reddened by dust, and similar trends are
observed in the red sequence at $z\sim 0$ \citep{bal06}. In the BzK sample, 11
red sBzKs with $SSFR<0.01^{-2}~Gyr^{-1}$ are all massive ($M > 10^{10}M_{\odot}$)
and generally red (rest-frame $U-V>1.3$ except two galaxies with $U-V>1.0$).
They are visually characterized by compact structures, with the exception of
one having large size ($R_{e}=3.5$ kpc) and a light profile characteristic of
an exponential disk ($n=1.34$). These could be passive galaxies that are not
classified as such by the BzK criterion either because of photometric scatter
in the photometric measures or because of the galaxies' intrinsic SED
variations (see Section 2.2).

We find 70 (50 for the BzK sample) X-ray detected galaxies among our sample,
marked as diamond symbols in Figure~\ref{fig:CM}. Most of them ($86\%$ :SED
sample, $90\%$: BzK sample) are star-forming galaxies (sBzKs), and those X-ray
detected galaxies are generally massive and compact. We do not exclude them
from further study since they also follow a similar trend in the color-mass
diagram, and have similar morphologies as non X-ray detected galaxies.

\section{Morphological Classification Using Non-parametric Approaches}

In order to investigate further the morphologies of galaxies within $1.4 < z
\le 2.5$, we turn next to several non-parametric morphological diagnostics
such as the Gini parameter ($G$), the second-order moment of the brightest
20\% of the galaxy pixels ($M_{20}$) and the multiplicity parameter ($\Psi$).
Many studies have used these parameters to quantify galaxy morphology
\citep{lot04, abr07, law07, ove10, law12, wan12} locally and at high redshift,
generally showing that they are an effective and automated way to measure
galaxy morphologies for large samples. These parameters quantify the spatial
distribution of galaxy flux among the pixels, without assuming a particular
analytic function for the galaxy's light distribution. Thus they may be a
better characterization of the morhology of irregulars, as well as standard
Hubble-type galaxies (Lotz et al. 2004, Hereafter, LPM04).  Before measuring
these parameters, we need to identify the pixels belonging to each galaxy. For
each galaxy, we calculate the ``elliptical Petrosian radius'', $a_{p}$, which
is defined like the Petrosian radius \citep{pet76} but uses ellipses instead
of circles (LPM04). We use the segmentation map generated by Sextractor when
making the H-band detections \citep{guo13}, and use the ellipticities and
position of peak flux determined by Sextractor for each galaxy. We then set
the semi-major axis $a_{p}$ to the value where the ratio of the surface
brightness at $a_{p}$ to the mean surface brightness within $a_{p}$ is equal
to 0.2. The surface brightness at each elliptical aperture, $a$, is measured
as the mean surface brightness within an elliptical annulus from $0.8a$ to
$1.2a$. There are 10 galaxies in the SED sample and one sBzK whose images
comprise less than 28 pixels (corresponding to a circle with a radius of 3
pixels), which we have excluded from further analysis. Note again that we use
galaxies with $M > 10^{9}M_{\odot}$ from both samples for our morphology
analysis, which leaves us with 46 pBzKs and 669 sBzKs, and 104 passive and
1567 star-forming galaxies of the SED sample.

Using the SED and BzK samples with stellar mass $>10^{9}M_{\odot}$, we first
compute the G parameter defined in LPM04, which measures the cumulative
distribution function of a galaxy's pixel values (light). Therefore, G of 1
would mean all light is in one pixel while G of 0 would mean every pixel has
an equal share. Hence, G is used to distinguish between the galaxies for which
fluxes are concentrated within a small region or uniformly diffuse. We also
compute the $M_{20}$ parameter, which traces the spatial distribution of any
bright nuclei, bars, spiral arms, and off--center star clusters.  Typically,
galaxies with high values ($M_{20} \ge -1.1$) are extended objects with double
or multiple nuclei, whereas low values ($M_{20} \le -1.6$) are relatively
smooth with a bright nucleus (see LMP04 for a detailed explanation of
$M_{20}$).  The third non-parametric coefficient is the multiplicity $\Psi$
\citep{law07}, designed to discriminate between sources based on how
"multiple'' the source appears.  Galaxies with lower $\Psi$ are compact
galaxies with generally one nuclei while irregular galaxies with multiple
clumps have higher $\Psi$ \citep{law07} (the definitions of each diagnostic
are presented in the above references).
\begin{figure*}
\begin{center}
\epsscale{0.8}
\plotone{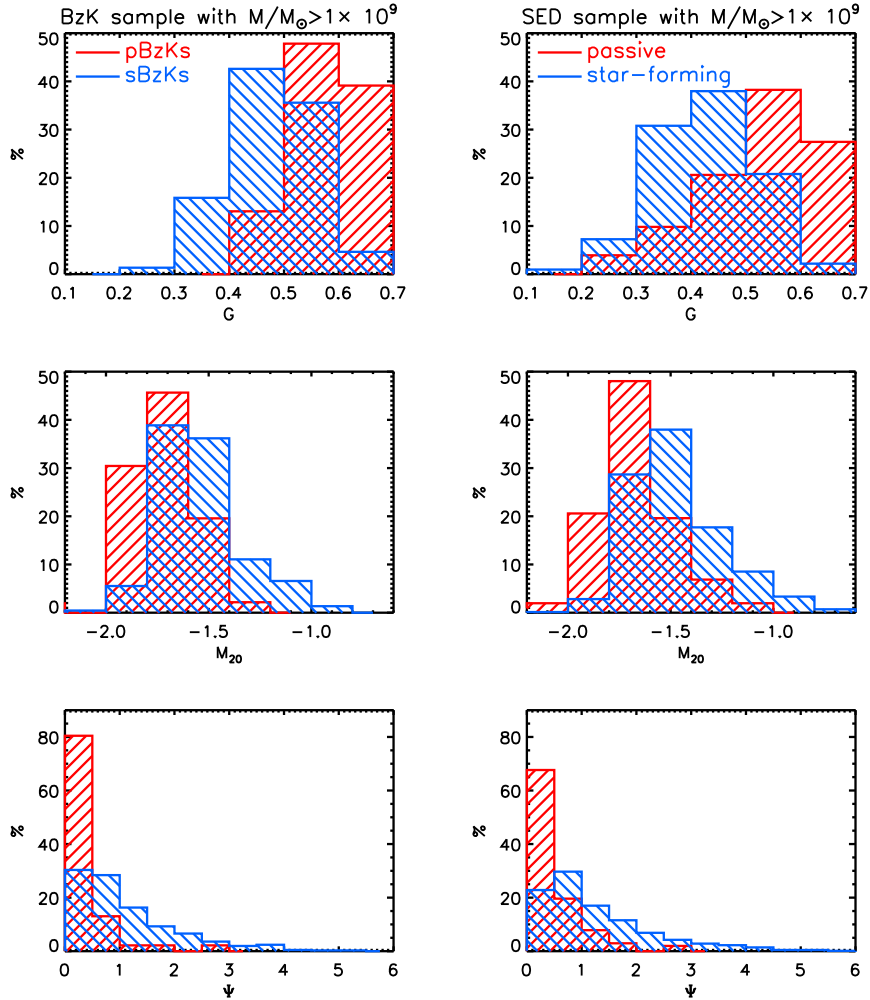}
\caption{\small Relative distributions of $G$ (top), $M_{20}$ (middle) and $\Psi$ 
(bottom) for BzK (left) and SED samples (right) having $M > 10^{9} M_{\odot}$. 
Red and blue histograms represent the pBzK (passive) and sBzK (star-forming) 
galaxies, respectively. Overall, both samples show similar morphological distribution 
in three parameters, such as passive (pBzK) galaxies have higher $G$ and lower 
$M_{20}$ and $\Psi$ in contrast to star-forming (sBzK) galaxies. This is consistent 
with the passive galaxies being compact and relatively smooth, while the star-forming 
ones are more extended and have more fine-scale structures.}
\label{fig:nonpar}
\end{center}
\end{figure*}

\subsection{Rest-frame Optical Morphology}

\begin{figure*}
\begin{center}
\epsscale{0.8}
\plotone{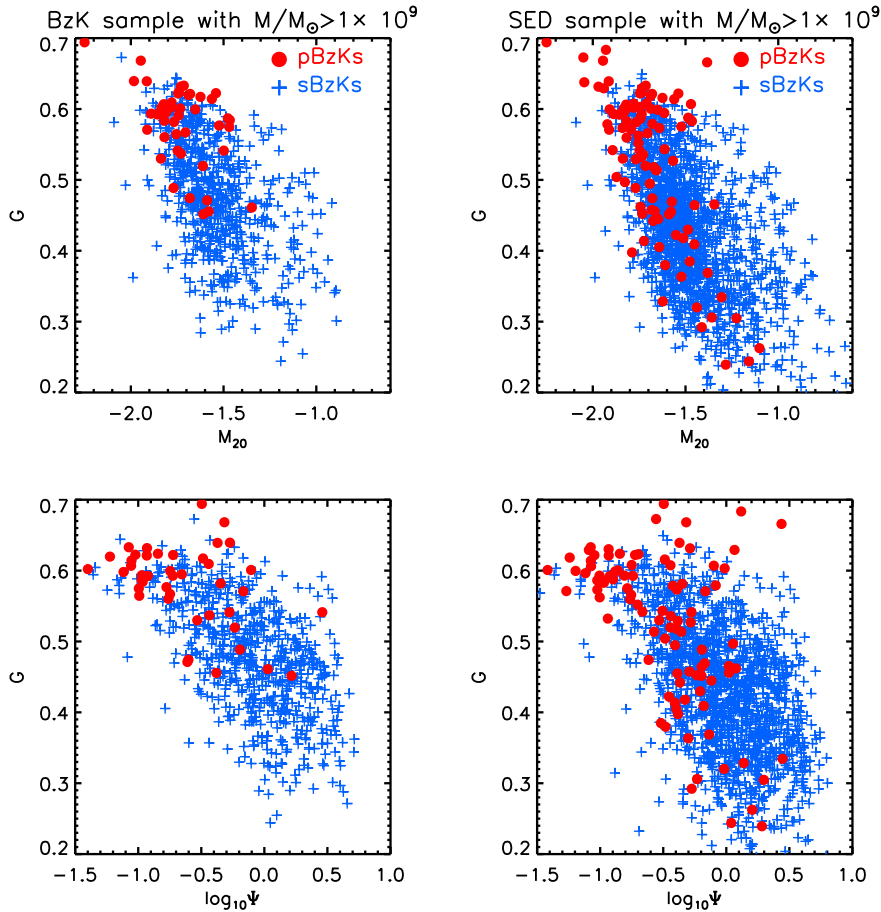}
\caption{\small Distribution of G vs. $M_{20}$ (top) and $\Psi$ (bottom) for BzK (left) and 
SED (right) samples having $M > 10^{9} M_{\odot}$. 
In both samples, there are clear morphological differences between passive galaxies 
and star-forming ones also seen in figure [\ref{fig:nonpar}]. However, in $G-M_{20}$ 
and $G-\Psi$ spaces, some star-forming galaxies show a similar morphological trend 
as passive one and several passive galaxies in the SED sample have lower $G$ and 
higher $M_{20}$ and $\Psi$ like star-forming ones. }
\label{fig:dist}
\end{center}
\end{figure*}

Figure~\ref{fig:nonpar} shows the relative distribution of the $G$, $M_{20}$
and $\Psi$ for star--forming and passive galaxies of the BzK and SED samples
(blue and red histogram, respectively). The $G$ values are mostly in the range
$0.3-0.7$ with a mean of 0.43 for star-forming galaxies (0.48 for sBzKs) and
0.53 for passive ones (0.58 for pBzKs). Passive galaxies are shifted to higher
$G$ than star-forming ones. The majority of pBzKs (90\%) and about 70\% of
passive galaxies have $G>0.5$. The mean values of the $M_{20}$ for
star-forming (sBzK) and passive (pBzK) galaxies are -1.47 (-1.54) and -1.68
(-1.73), respectively. The middle panel of Figure~\ref{fig:nonpar} shows that
the passive galaxies (pBzKs) have lower values and show a peak at $\sim -1.7$
while the star-forming ones (sBzKs) exist in a wide range of $M_{20}$ values
that are slightly skewed to higher $M_{20}$.  Lastly, the $\Psi$ values of
star-forming galaxies (sBzKs) have a range of values up to $\sim 5$, but most
of the passive galaxies (SED: 90\% , BzK: 94\% ) have $\Psi <
1.0$. \cite{law12} find that spectroscopically confirmed star-forming galaxies
at z=1.5--3.6 have $\Psi < 1$ for isolated regular galaxies, $1 < \Psi < 2$
for sources that show some morphological irregularities, and larger values for
sources having multiple clumps that are separated. Therefore, all passive
galaxies (pBzKs) tend to be dominated by one main clump while star-forming
ones (sBzKs) can have two or more significant components in addition to a main
nucleation.  There is some degree of correlation between the $G$ and $M_{20}$,
$\Psi$ measurements (see Figure~\ref{fig:dist}). The passive galaxies (pBzKs)
reside in a narrow region with higher $G$ and lower $M_{20}$ and $\Psi$
indicating that they consist of one bright central source. In contrast with
passive galaxies (pBzKs), star-forming ones (sBzKs) with lower $G$ have higher
$M_{20}$ and $\Psi$ because galaxies with diffuse morphology tend to have a
spread out flux distribution. Figure~\ref{fig:nonpar} and \ref{fig:dist} also
show that there is an overlap in the distributions of morphological parameters
of star-forming galaxies and passive ones. For example, there are star-forming
galaxies with high $G$ and low $\Psi$ or $M_{20}$ that are located in same
region where the bulk of passive galaxies are observed, and vice versa.

\begin{figure}
\begin{center}
\epsscale{0.9}
\plotone{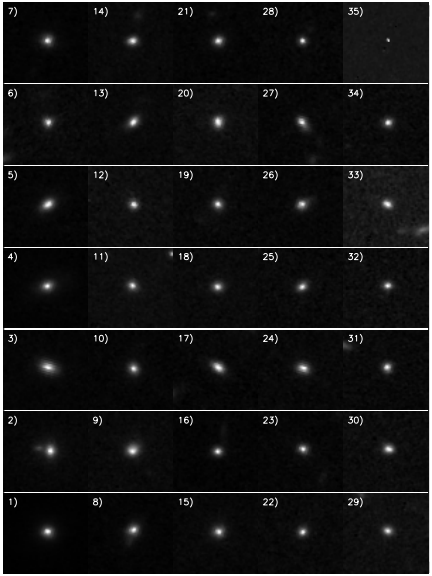}
\caption{ \small Postage stamps of 35 star-forming galaxies ( in SED sample) with $G > 0.6$. 
Each postage stamp is $3.6 \times 3.6~arcsec^2$ and all images have been linearly scaled. 
The number in each stamp indicates the order of H-band magnitude, i.e. number 1 galaxy is the 
brightest one. As one can see, all star-forming galaxies with high $G$ show spheroid-like 
structures with a bright clump.}
\label{fig:sfg}
\end{center}
\end{figure}

\begin{figure}
\begin{center}
\epsscale{0.8}
\plotone{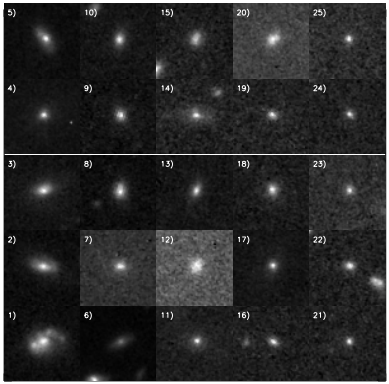}
\caption{\small Postage stamps of 25 passive galaxies ( in SED sample) with $G < 0.5$ 
(Image stamp size, image scaling and magnitude order follow the same properties of Figure~\ref{fig:sfg}). 
Among 35 passive galaxies with $G>0.5$, 10 galaxies are excluded due to 
the low signal-to-noise ratio per pixel, $S/N_{p.p} < 2.5$. Those galaxies 
are all red and show extended stuctures as an example of red (passive) disks.}
\label{fig:pg}
\end{center}
\end{figure}

To illustrate and further explore these galaxies in the overlapped region, we 
have chosen star-forming galaxies with $G> 0.6$ and passive
galaxies with $G<0.5$ for visual inspection and classification. We indeed
found that the star--forming galaxies can generally be classified as blue
spheroids and the passive ones as red disks. In agreement with \cite{law07},
the 35 star-forming galaxies with high $G$ visually appear as compact structures
in Figure~\ref{fig:sfg}. Note that all these images have S/N ratio per pixel
($S/N_{pp}$) greater than 2.5, the threshold used in LPM04 for reliable
measurements, and most of them (85\%) are relatively bluer than normal passive
galaxies. About 40\% of star-forming galaxies in the sample of \cite{law12}
were visually classified as such compact structures as well.  Among passive
galaxies in the SED sample, 10 of them have $S/N_{pp}<2.5$, and all have
$G<0.5$. This indicates that we cannot measure reliable morphology due to the
low $S/N_{pp}$ ratio.  In Figure~\ref{fig:pg}, we show the 25 passive
galaxies with $G<0.5$ and $S/N_{pp} >2.5$. Most of them have smooth
structures, and some are elongated or have secondary structures. They are
intrinsically red with rest-frame $(U-V) > 1.5$, and 16 galaxies are massive
($M>10^{10}M_{\odot}$). Red (passive) disks at high redshift have also been
recently studied by other groups. For example, \cite{wan12} found that 30\% in
quiescent galaxies of their sample with $M>10^{11} M_{\odot}$ at $1.5\le z \le
2.5$ can be morphologically classified as disks. This is generally consistent
with the findings presented here, although we note that due to their low
$S/N_{pp}$, some of our ``passive disks'' might actually be morphological
mis-classifications or even be dust--obscured star--forming galaxies. Finally,
we observe that, overall, the BzK and the SED samples have essentially
identical distributions of morphological parameters, although the SED sample
includes galaxies with lower $G$, namely those with $S/N_{pp} < 2.5$, which
is the result of their lower surface brightness.

\begin{figure}
\begin{center}
\epsscale{1.1}
\plotone{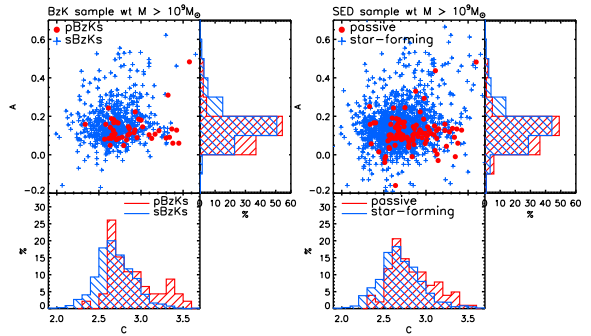}
\caption{\small The plot of Asymmetry[A] vs. Concentration[C] for BzK(left) and SED (right) 
samples having $M > 10^{9} M_{\odot}$ and the histograms of each parameter. 
The passive galaxies (pBzKs) are more spheroidal like in their C 
and A values, while star-forming galaxies (sBzKs) are more spiral and 
merger like. Passive galaxies (pBzKs) are different in C-A plane than the star-forming 
ones (sBzKs) although the difference is not huge compared to $G-M_{20}$ and $-\Psi$.}
\label{fig:cas}
\end{center}
\end{figure}

In addition to $G$, $M_{20}$ and $\Psi$, we measure the Concentration ($C$)
and Asymmetry ($A$) of our samples. The concentration index $C$
\citep{abr96,con03} measures the concentration of flux. Typical values of $C$
range from $\sim 1$ for the least compact to $\sim5$ for most compact
galaxies.  Note that asymmetry $A$ \citep{sch95, abr96, con00} compute the
$180$ degree rotational asymmetric light distribution of all galaxy and hence
the most symmetric galaxies have $A=0$.  We present the distribution of $C$
and $A$ for the BzK (left) and SED (right) samples in Figure~\ref{fig:cas}. As
expected, the passive galaxies (pBzKs) are more similar to ellipticals in
their $C$ and $A$ values, while the star-forming ones (sBzKs) are more spiral
and merger-like. In the $C-A$ plane, passive galaxies (pBzKs) are different
from star-forming ones (sBzKs), but the difference is not as significant
compared to the difference in $G-M_{20}$ and $\Psi$.  As expected, the
distribution of $C$ and $A$ for both samples are similar as shown in the
previous figures.

In summary, the distributions of the non--parametric morphological diagnostics
that we have considered here for both the BzK and SED samples are essentially
the same in each spectral type class. Star--forming and passive galaxies
clearly show different distributions of non-parametric morphological measures,
and they are separated well in $G$--$M_{20}$ and $\Psi$ spaces. Passive
galaxies (pBzKs) are mostly compact, spheroidal structures, and the majority
of star-forming ones (sBzKs) are somewhat extended or have multiple clumps,
similar to disks or irregular galaxies in the Local universe. These results
agree with those of \cite{wan12} who also studied the morphologies of massive
galaxies ($M > 10^{11}M_{\odot}$) at $z\sim2$ with $G$, $M_{20}$ and visual
classifications. They found that the quiescent galaxies are bulge dominated
and star-forming galaxies have disks or irregular morphologies visually as
well as in the $G$ and $M_{20}$ analysis. We extend their study with a larger
sample down to a lower mass limit, and obtain almost the same conclusion about
galaxy morphologies at $z\sim 2$.
\begin{figure}
\begin{center}
\epsscale{1.0}
\plotone{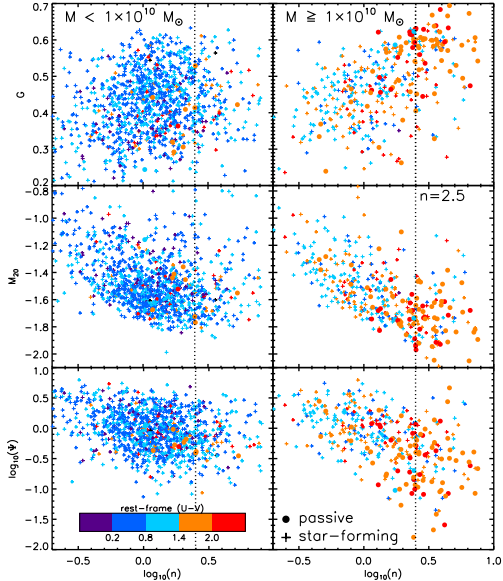}
\caption {\small The distribution of morphological parameters as a function of 
$S\acute{e}rsic$ index ($n$) for the SED sample with $M> 10^{9}M_{\odot}$ in two 
different mass bins divided by a mass threshold, $M_{th}=10^{10}M_{\odot}$. 
The color-coding represents the rest-frame (U-V) color of galaxies, and the 
dotted vertical line is for $n=2.5$. About 96\% passive galaxies (circles) are more 
massive than $M_{th}$, while about 78\% of star-forming ones (crosses) have mass, 
$M < M_{th}$.  Star-forming galaxies dominated by low $n$, especially at $M<M_{th}$ 
and passive ones mostly have higher $n$. Redder galaxies (mostly passive galaxies) tend to 
have higher $G$ and lower $M_{20}$ and $\Psi$ in the massive system.}
\label{fig:sersic}
\end{center}
\end{figure}

\begin{figure}
\begin{center}
\epsscale{1.0}
\plotone{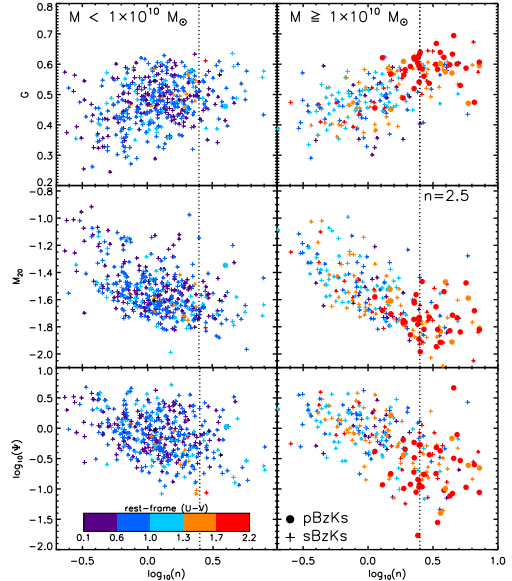}
\caption {\small The distribution of morphological parameters as a function of $S\acute{e}rsic$ index 
($n$)  for the BzK sample with $M> 10^{9}M_{\odot}$ in two different mass bins divided 
by a mass threshold, $M_{th}= 10^{10}M_{\odot}$. The color-coding represents 
the rest-frame (U-V) color of BzK sample, and the dotted vertical line is for $n=2.5$. 
All pBzKs (circles) are more massive than $M_{th}$, while about 70\% of sBzKs (crosses) 
have mass, $M < M_{th}$. sBzKs dominated by low $n$ and pBzKs mostly have higher 
$n$. The morphologies and rest-frame colors are well separated, especially in the massive systems. } 
\label{fig:sersicb}
\end{center}
\end{figure}

In Appendix A, we investigate the robustness of non-parametric measures ($G$,
$M_{20}$ and $\Psi$), mainly used in this study for morphological analysis,
using GOODS-S and the Hubble Ultra Deep Field (UDF) images in the H-band. The
UDF overlaps part of the GOODS-S imaging, but goes much deeper (5 $\sigma$
depth of 28.8), and thus offers an opportunity to test the dependence of
parameters on the signal-to-noise per pixel ($S/N_{pp}$). We show that any
difference between the two different fields, which have different exposure
times, is relatively small for three parameters, with the scatter in measured
properties increasing as S/N decrease.  We find that most ($> 90\%$) of BzK
galaxies have $S/N_{pp} > 2.5$, and $\sim70\%$ of the SED sample have
$S/N_{pp} > 2.5$.  We note that we do not exclude galaxies with $S/N_{pp} <
2.5$ since they rarely change our results in this study.

\subsection{$G$, $M_{20}$ and $\Psi$ vs. $S\acute{e}rsic$ Index and $R_{e}$}

\begin{figure}
\epsscale{1.0}
\plotone{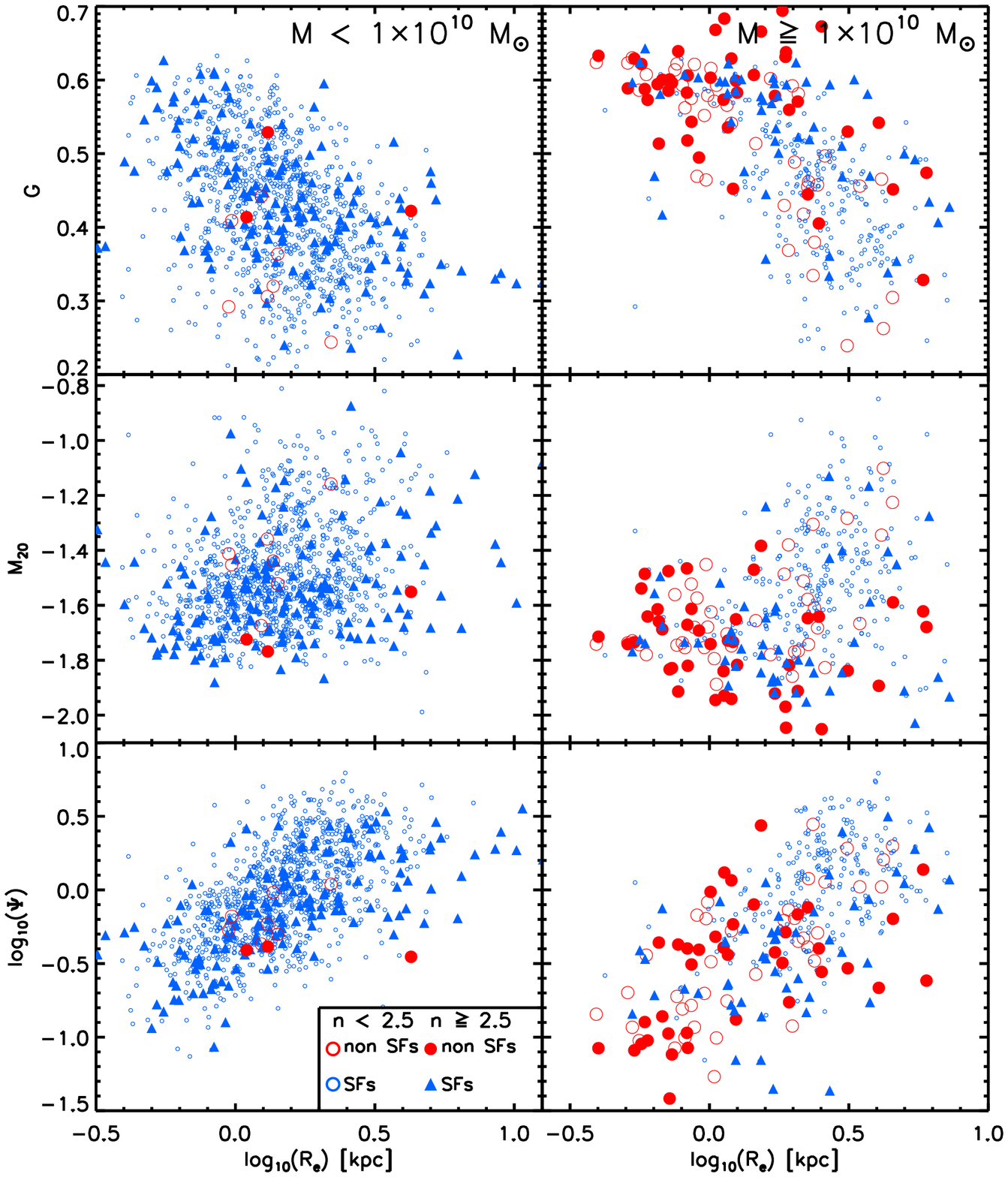}
\caption{\small The distribution of morphological parameters as a function of half-light 
radius ($R_{e}$) for the SED sample with $M> 10^{9}M_{\odot}$ in two different 
mass bins divided by a mass threshold, $M_{th}= 10^{10}M_{\odot}$. Open symbols 
show the galaxies with $n < 2.5$ and filled symbols show galaxies with $n > 2.5$. 
Star-forming and passive galaxies are expressed as blue and red colors, respectively. 
Overall, star-forming galaxies tend to have larger effective radii than passive ones, 
even in the case of massive systems, $M > M_{th}$, and a half of passive galaxies 
show very compact morphologies, with $R_{e} < 1 kpc$. Galaxies with smaller 
$R_{e}$ tend to have a compact structure with high $G$ and low $\Psi$. }
\label{fig:re}
\end{figure}

\begin{figure}
\epsscale{1.0}
\plotone{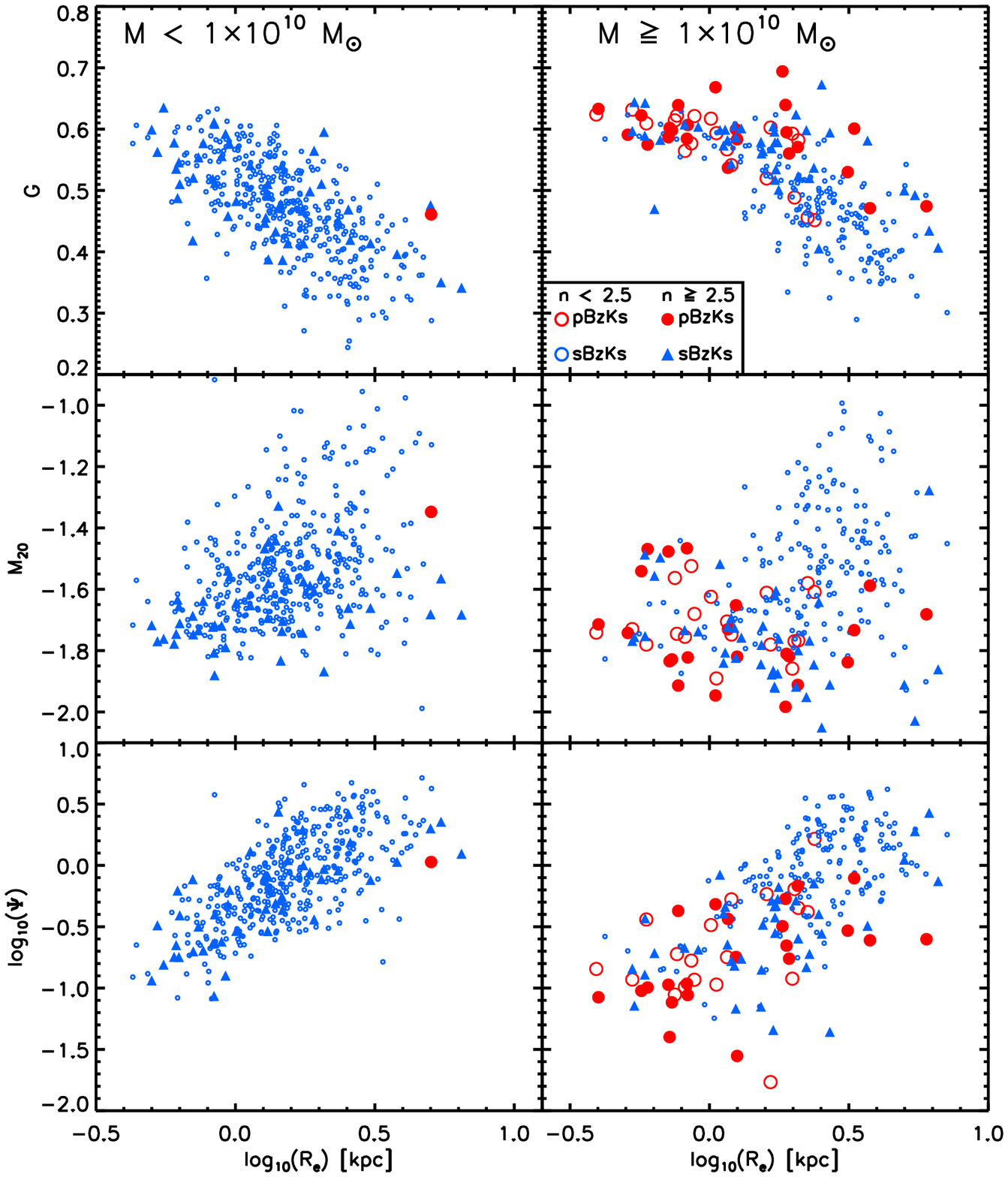}
\caption{\small The distribution of morphological parameters as a function of half-light 
radius ($R_{e}$) for the BzK sample with $M> 10^{9}M_{\odot}$ in two different 
mass bins divided by a mass threshold, $M_{th}=10^{10}M_{\odot}$. 
Open symbols show the galaxies with $n < 2.5$ and filled symbols show 
galaxies with $n > 2.5$. sBzKs and pBzKs are expressed as blue and red colors, 
respectively. The statistics of $R_{e}$ and overall morphologiacl distributions 
are same with SED sample of Figure~\ref{fig:re}.} 
\label{fig:reb}
\end{figure}

$S\acute{e}rsic$ index and half-light radius ($R_{e}$) have been successfully
used to characterize galaxy morphology in many previous works, both at low and
high redshift (low-z: Blanton et al. 2003, Driver et al. 2006; mid-z: Cheung
et al. 2012; high-z: Ravindranath et al. 2006, Bell et al. 2012). Recently,
\cite{bel12} and \cite{wuy11} show that the $S\acute{e}rsic$ index correlates
well with quiescence in galaxies at $z \lesssim 2$. Therefore, we investigate
how galaxy morphologies with $G,M_{20}$ and $\Psi$ correlate with
$S\acute{e}rsic$ index (n) and $R_{e}$. We use the $S\acute{e}rsic$ index and
$R_{e}$ \citep{vdw12} obtained by fitting a $S\acute{e}rsic$ profile to the
galaxy image using GALFIT \citep{pen02}. Passive galaxies in both samples have
$\langle n\rangle\sim 3.0$, and over 96\% of them have $n>1.0$, with 50\%
having $n>2.5$. In contrast, star--forming galaxies have $\langle n\rangle\sim
1.5$, with 85\% of them having $n<2.5$. This suggests that the majority of
star-forming (sBzK) galaxies have disk-like (exponential light profile) or
irregular structure with a light profile shallower than an exponential one.
In contrast, all passive galaxies (pBzKs) have a dominant bulge including some
bulge+disk structures. A similar analysis of morphologies at $z<2.5$ using the
$S\acute{e}rsic$ index in the SFR-mass diagram was carried out by
\cite{wuy11}, who found that the main sequence (MS) consists of star-forming
galaxies with near exponential profiles, and passive galaxies below the MS
have higher $S\acute{e}rsic$ indices close to a de Vaucouleur profile
($n=4$). \cite{szo11} also reported similar results with 16 massive galaxies
at $z\sim2$, and found that star-forming galaxies have diskier (low n)
profiles than passive galaxies. We present the distribution of $G,M_{20}$ and
$\Psi$ as a function of $S\acute{e}rsic$ index in two different mass bins
divided by a threshold mass, $M_{th}=1\times10^{10}M_{\odot}$ in
Figure~\ref{fig:sersic} (SED sample) and ~\ref{fig:sersicb} (BzK sample).

In both samples, we find that there are significant correlations between
$S\acute{e}rsic$ index and $G,M_{20}$ and $\Psi$, with galaxies with high $n$
having high $G$ and low $M_{20}, \Psi$, and vice versa. As we have already
noted in Figure~\ref{fig:CM}, most of the passive galaxies (pBzKs) have masses
greater than $1\times10^{10}M_{\odot}$, and the majority of star-forming
galaxies (78\%) and $70\%$ for sBzKs have $M < 1\times10^{10}M_{\odot}$. In
massive systems ($M \ge 1\times10^{10}M_{\odot}$), the two populations show a
well-separated bimodal distribution in their morphologies and colors (see
Figures 10 and 11). Red passive galaxies (pBzKs) show spheroidal-like
structures with high $n$, $G$ and low $M_{20}$, $\Psi$, while blue
star-forming ones (sBzKs) exhibit a larger variety of morphologies, but mainly
have low $n$, $G$ and high $M_{20}$, $\Psi$. There are some star-forming
galaxies (sBzKs) with high $S\acute{e}rsic$ index ($n > 2.5$, vertical dotted
line in Figure~\ref{fig:sersic},~\ref{fig:sersicb}). They follow mostly the
same trend in morphologies with higher $G$ and lower $M_{20}$, $\Psi$,
indicating the presence of a bright center (see the blue spheroids in
Figure~\ref{fig:sfg}).  \cite{bel12} showed examples of such systems in their
sample, and found that those appear to be spheroidal-like structures, but in
many cases also have significant asymmetries, or faint secondary sources and
tidal tails. A loose relation between non-parametric measures and
$S\acute{e}rsic$ index is observed for massive galaxies in the right panel of
Figure~\ref{fig:sersic}, \ref{fig:sersicb}, but not for low mass galaxies at
all.  This means that the commonly used $S\acute{e}rsic$ index is not enough
to study morphology of those galaxies. Therefore, it is important to use the
non-parametric diagnostics in addition to the $S\acute{e}rsic$ profile to
quantify the morphology of galaxies towards the low end of the mass
distribution.

In Figure~\ref{fig:re} (SED sample) and ~\ref{fig:reb} (BzK sample), we plot
the non-parametric measures as a function of $R_{e}$ in both small and large
mass systems to see how the size varies along with spectral type and stellar
masses. We find that $R_{e}$ also correlates well with all the non-parametric
measures in general, as galaxies with low $G$ and high $M_{20}$, $\Psi$ have
smaller sizes and relatively low $S\acute{e}rsic$ Index ($n<2.5$: Empty
symbols) in both samples. Overall, star-forming galaxies (sBzKs) tend to have
larger half-light radii than passive ones (pBzKs), even in the case of massive
systems ($M \geq 1\times 10^{10}M_{\odot}$) and about half of passive (pBzK)
galaxies show very compact morphologies, with $r_{e} < 1~kpc$. This is
consistent with previous results, which find that passive galaxies are more
compact than star--forming galaxies at $z\sim 2$ \citep{tof09, wuy11, cas11},
and the same general trend is observed at $z\sim 0$ among massive galaxies
\citep{wil10}.

\section{Comparison with the Local Universe}
\begin{figure*}
\epsscale{1.0}
\plotone{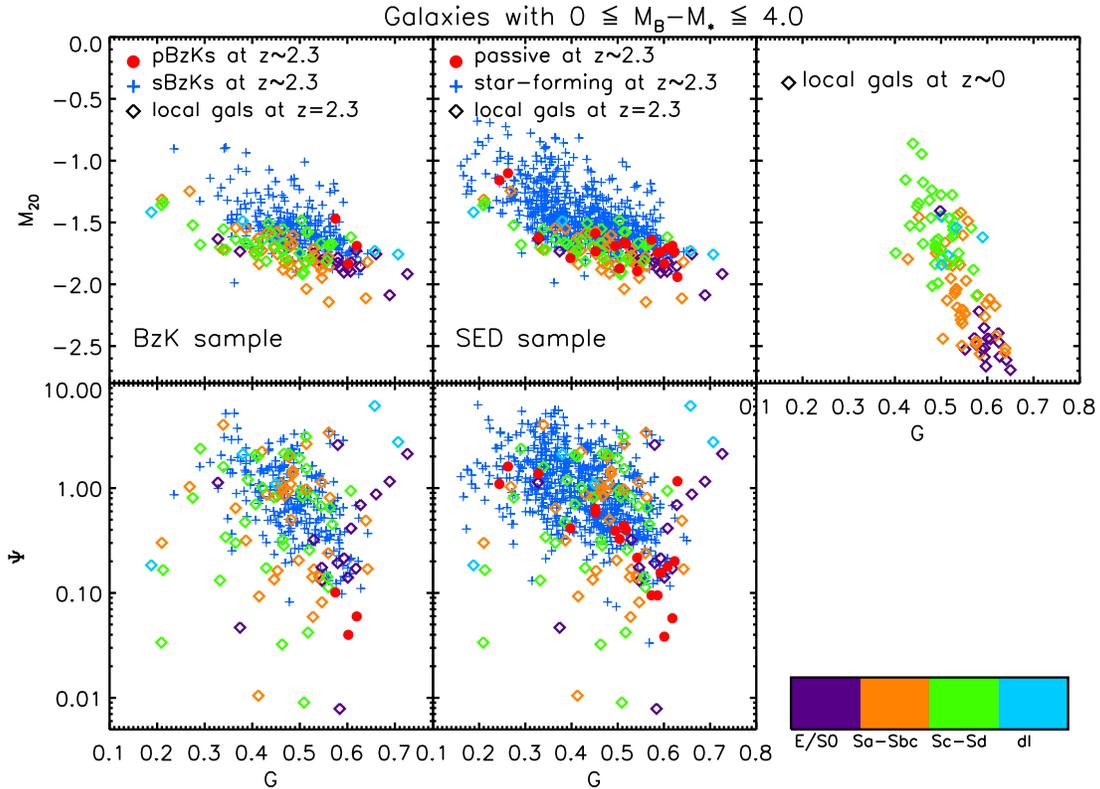}
\caption{\small $G$ vs. $M_{20}$ and $\Psi$ for the galaxies at $z\sim2.3$ and local galaxies from LPM04. 
We compare the morphologies of BzK [from left: 1st panel] and SED sample [2nd panel] in
 WFC3 H-band with degraded B/g-band image of local galaxies. To reduce the effect of 
any morphological k-correction, we compare only the 
galaxies with redshifts $ 2.0 < z \le 2.5$. Local galaxies are selected to lie in the 
same $M_{B}-M*$ range ($0.0 \le (M_{B}-M*) \le 4.0$) as the $z\sim 2.3$ sample, 
assuming $M*=-20.1$ locally and $M*=-22.9$ at $z\ge 2$. 
The 3rd panel shows observed morphologies of normal local galaxy types expressed 
by the following colors ( violet: E/S0, magenta: Sa-Sbc, green: Sc-Sd, light blue: dI). 
Comparison between passive and star-forming galaxies in each sample 
(red dots and blue crosses, respectively) at $2.0 < z \le 2.5$ and the morphologies of 
redshifted local galaxies at the WFC3 H-band image resolution are described 
on the 1st and 2nd panels. Overall, galaxies at $z\sim2.3$ tend to have similar distribution 
in $G-M_{20}$ space with redshifted local galaxies even though there are many galaxies with higher 
$M_{20}$ for their $G$ than for any of the local ones. }
\label{fig:local}
\end{figure*}

The strong correlation between galaxy color (and SFR) and morphology shown in
the previous sections is reminiscent of the Hubble sequence at $z=0$. However,
to understand if actually the Hubble sequence is in place at $z\sim2$, it is
important to examine how galaxy morphologies at $z\sim2$ differ from the local
galaxies. In general, comparing morphological parameters of local galaxies,
which are observed at relatively high resolutions to their conterparts in
high-z samples, whose resolution is lower, is not straightforward because
because most morphological diagnostics do depend on the resolution. A
fortunate case, however, is that of the comparison between local galaxies at
redshift $0.05<z<0.1$ from the SDSS survey to galaxies at $z\sim 1$ observed
with {\it HST}, since the difference of angular diameter distance at these two
redshifts nearly perfectly compensates for the difference in the angular
resolution of the two instrumental configuration \citep{nai10}. At this
purpose, we should also keep in mind that in our adopted cosmology, the
fractional variation of the angular diameter distance in the redshift range
$1.4<z<2.5$ is only $\approx 5$\%. Furthermore, even though $G$ and $M_{20}$
measures are robust given the resolution of the observation, particularly if
the data is deep enough to allow the Petrosian radius to be used to measure
the parameters \citep{abr07}, we should nonetheless be careful when directly
comparing the $G$ and $M_{20}$ from observations with different resolutions
(LPM04; Lisker et al. 2008).  Therefore, in this study, we compare the
$z\sim2$ galaxy morphologies in the SED and BzK samples to those of the local
galaxy sample of LPM04 after we simulate how they would appear in the CANDELS
images if they were observed at redshift $z\sim 2$. For this reason, we have
used the B--band and g--band images of the local galaxies \citep{fre96,
  aba03}, which at this redshift correspond to the H band (for details about
the local galaxy observations, see LPM04). We have restricted the comparative
analysis with the high--redshift galaxies to only those at $2.0 < z \le 2.5$
to minimize the possible effects of the morphological
K-correction. Furthermore, we have only considered galaxies within a magnitude
range of $0\le(M_{B}-M*)\le4$ (LPM04), where we take $M*=-20.1$ \citep{bla03b}
for local galaxies, and $M*=-22.9$ for galaxies at $z \ge 2$ \citep{sha01},
assuming that the local galaxies were brighter in the past but did not evolve
morphologically (LPM04).

In this simulation, we first modify the angular sizes and surface brightness
of local galaxy images to account for distance and cosmological effects. The
images are rebinned to the pixel scale of the galaxies observed at $z=2.3$
(WFC3 pixel size is 0.06") and the flux in each pixel is rescaled so that the
total magnitude of the galaxy corresponds to some preassigned value, for
example to that of an M* galaxy at $z=2.3$. The modified images are then
convolved with the WFC3 PSF and, lastly, we add Poisson noise appropriate to
the WFC3/NIR observations using the IRAF task MKNOISE.

In Figure~\ref{fig:local}, we present the $G$, $M_{20}$ and $\Psi$ measured
from the redshifted modified local galaxy images. The measures of the
redshifted galaxies (from the left, 1st and 2nd panels) are quite different
from those of the original local galaxy images (3rd panel). This is in
agreement with LPM04, who conclude that $z\sim2$ Lyman Break Galaxies (LBGs)
do not have morphologies identical to local galaxies. Overall, the
distributions of galaxies at $z\sim 2.3$ in both samples and that of the
redshifted local galaxies are similar in $G$-$M_{20}$ space, as shown in the
1st and 2nd panel of Figure~\ref{fig:local}, but the high--redshift
star--forming galaxies have a broader distribution of $M_{20}$ for a given
value of $G$ than the redshifted local late types. This trend is reflected in
the $G$-$\Psi$ plane (in the bottom panels), which shows lower $\Psi$ values
for the redshifted local galaxies. Large and luminous star--forming disks are
mostly responsible for this excess of galaxies with higher $M_{20}$ and
broader (slightly higher) distribution of $\Psi$, another manifestation of the
fact that disks at $z\sim 2$ are not simply scaled--up versions of the local
ones in terms of star--formation rates, but are intrinsically different
(e.g. Papovich et al.\ 2005, Law et al.\ 2007).  Comparatively, the $z\sim 2$
passive galaxies have $G$ and $M_{20}$ values that are much more similar to
those observed for the present--day E/S0 galaxies. Overall, the qualitative
similarity of values, shapes and trends of the distributions of morphological
parameters at low and high redshift suggests that the Hubble sequence is
essentially in place at $z\sim 2$. 

Our comparative study also shows that while the morphology of the oldest
systems at any epoch, i.e. the passively evolving galaxies, in general changes
relatively little from $z\sim 0$ to the present, at least as traced by our
diagnostics, disk galaxies underwent strong structural evolution over the same
cosmic period. A noticeable exception is the evolution of the size of massive
ellipticals, which at $z\sim 2$ were dominated by very compact galaxies, which
had stellar density up to two order magnitudes higher than today's
counterparts of similar mass, while at present such systems have essentially
disappeared (see \cite{cas11, cas13}).  Also, it is interesting to note that
\cite{hua13} find that at even higher redshifts, i.e. $4<z<5$, the size
distribution of star--forming galaxies is significantly larger than that
predicted from the spin parameter distribution observed in cosmological
N--body simulations, a marked difference from local disks which follow the
simulation predictions very well.

\section{ Comparison With Rest-Frame UV}

\begin{figure}
\centering
\includegraphics[width=9cm,height=16cm]{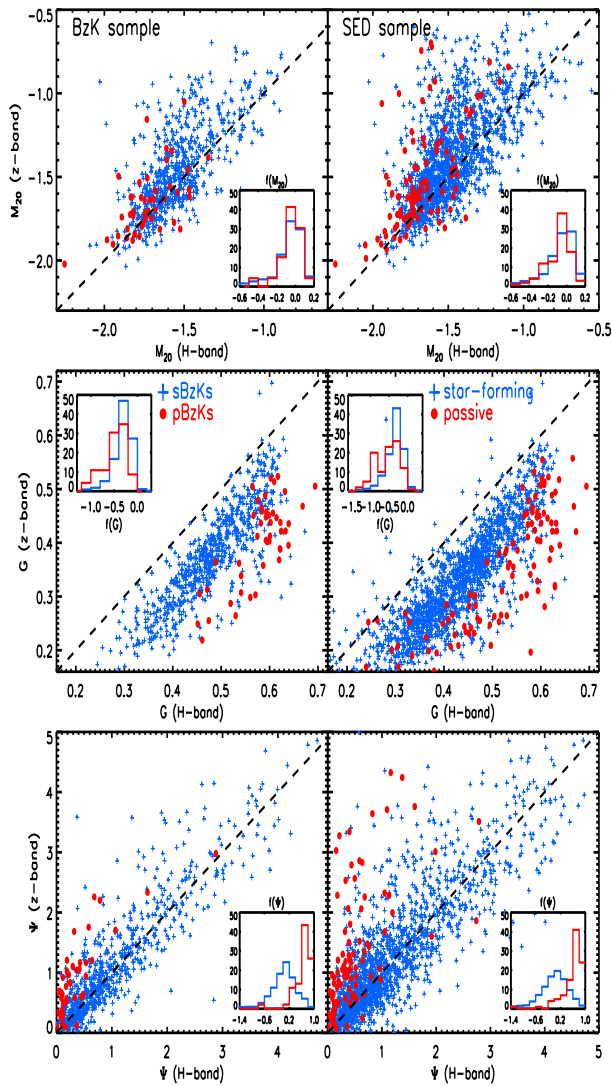}
\caption{\small Comparison of the morphological parameters for galaxies at $z\sim2$ 
between the rest-frame UV (z-band) and optical (H-band) [BzK: left, SED: right]. 
The top, middle and bottom panels show $M_{20}$, $G$ and $\Psi$, respectively. 
Blue crosses and red circles represent the sBzKs (star-forming) and pBzKs (passive), 
respectively, and a dotted black line in each panel shows a linear correlation. 
The inset in each panel show the distribution of the fractional differences ($f$) 
of the parameters in the two rest-frame bands defined as 
$f(M_{20})=[M_{20}(z)-M_{20}(H)]/M_{20}(z)$. Negative $f(M_{20})$ and 
positive $f(G),~f(\Psi)$ means that parameters are bigger in the z-band in each plot.}
\label{fig:zh}
\end{figure}

It is important to compare the rest--frame UV and optical morphologies, since
the former traces the spatial distribution of star formation, and thus
contains information on how, and where, galaxies grow in mass and evolve
morphologically, while the latter traces the structure of their stellar
components.  With the CANDELS images we can study the relation between
rest-frame optical (5300~\AA) and UV (2800~\AA) morphology with a large
sample. Since non-parametric measures can vary systematically with the PSF and
pixel size of the images as the resolution decreases (Lotz et al. 2004; Law et
al. 2012), we made a version of the ACS z--band image which is PSF--matched to
the WFC3 H--band one (by convolution with an ad--hoc kernel) and which we have
re-binned to the same pixel scale of 0.06"/pixel. To make a meaningful
comparison, we use the same segmentation map and the ``elliptical Petrosian
radius" estimated from the H-band image and apply it to the PSF matched z-band
image to measure the $G,~M_{20}$ and $\Psi$ in the rest-frame UV. In
Figure~\ref{fig:zh} we compare the $G,M_{20}$ and $\Psi$ of the BzK and SED
samples in the UV and optical rest-frames (passive (pBzK) and star-forming
(sBzK) galaxies in red and blue, respectively). What we find from our
measurements is that all three parameters are different between the H- and
z-bands in general. First, the H-band derived $G$ is higher for both passive
(pBzKs) and star-forming (sBzKs) galaxies at $z\sim2$. The values of $M_{20}$ 
at the two wavelengths are well correlated for both populations, but generally
the z-band measurements have slightly higher values.  In particular, we find
that the scatter increases as $M_{20}$ increases. This means that galaxies are
clumpier in the z-band and the difference in $M_{20}$ between the two bands is
bigger in the case of multi-clump structures (e.g. higher $M_{20}$). The
$\Psi$ values from the H-band and z-band for star-forming (sBzK) galaxies are
well correlated, while most of the passive ones (pBzKs) in the z-band have
higher $\Psi$ value than in the H--band. We compute the fractional differences
of the three parameters between optical and UV defined as
$f(M_{20})=[M_{20}(z)-M_{20}(H)]/M_{20}(z)$, shown in the insets of
Figure~\ref{fig:zh} to check for offsets from the linear correlation. Negative
$M_{20}$, positive $G$ and $\Psi$ imply that parameters in the z-band are
higher than the ones in the H-band.

Overall, in the rest-frame UV, galaxies appear to have higher $\Psi$ and
$M_{20}$ and lower $G$ values than in the rest-frame optical since
observations in the rest-frame UV show more fragmented structures than the
rest-frame optical, especially for star-forming galaxies. We additionally find
that the passive galaxies (pBzKs) in the rest-frame optical tend to have a
higher $G$ and lower $\Psi$ and $M_{20}$ than in the rest-frame UV because the
rest--frame optical light from old stellar populations is typically more
concentrated than that from younger stellar populations, consistent with the
result from \cite{guo11} that the inner region of passive galaxies at $z\sim2$
have a redder color gradient. A similar trend was noted by \cite{wuy12}, who
found that the median galaxy size and $M_{20}$ are reduced (less clumpy),
while $G$ and $C$ increase from rest-frame 2800~\AA\ and U band to the optical,
using star-forming galaxies at $0.5 < z < 2.5$.
\begin{figure}
\centering
\includegraphics[width=9.5cm,height=13cm]{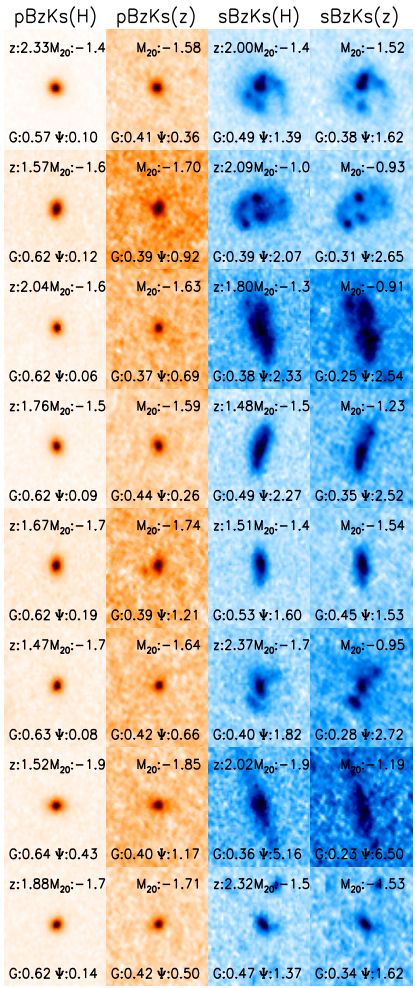}
\caption{\small Postage stamps of 16 galaxies which are selected in both BzK and SED samples, 
including 8 pBzKs (1st \& 2nd columns) and 8 sBzKs (3rd \&4th columns) in
the rest--frame optical (WFC3 H-band: 1st \& 3rd) and UV (ACS z-band: 2nd \& 4th). 
The galaxy images are ordered by decreasing magnitude from top to bottom 
(the eighth galaxy is the brightest one at each column). Each postage stamp is $3.6 \times 3.6~
arcsec^2$ and labels indicate the redshifts ($z$) and morphological parameters. 
Star-forming galaxies show a variety of morphologies, while all passive galaxies show bulge--like structures. 
The morphology between z and H-band of passive galaxies are almost identical, 
but this is not the case for star-forming ones. }
\label{fig:mont}
\end{figure}
Sample images of galaxies in both rest-frame optical and UV are shown in
Figure~\ref{fig:mont} to visually present the morphological differences
between passive and star-forming galaxies and to see how nonparametric
measures are correlated with the visual classifications. We have selected 16
galaxies (eight pBzKs and eight sBzKs) included in both BzK and SED
samples. These images have relatively smooth and regular morphologies in both
bands. The eight images for each spectral type are sorted by their H-band
magnitude, with the brightest one located at the bottom and decreasing upward,
and non-parametric statistics and redshifts of each galaxy labeled.  This
figure illustrates the good correspondence between the measured parameters and
the visual morphologies of the galaxies. Most sBzKs are extended, and exhibit
a broad range of morphologies, from isolated systems with a central bulge and
galaxies with bulge and disk components, to irregular systems with
multi-clumps. The pBzKs have morphologies similar to those of present day
spheroids, bulges and ellipticals. Visually, the sBzKs appear clumpier in the
rest-frame UV compared to optical, showing a dependence on wavelength with the
exception of a few isolated cases, while pBzKs look very similar in both
wavelengths. Overall, visual inspection shows that the morphological types in
both bands are generally similar, in agreement with \cite{cas10} who found
that passive galaxies ($SSFR <0.01~Gyr^{-1}$) have a ``weak morphological
K-correction", with size being smaller in the rest-frame optical than in the
UV. However, the comparison with non-parametric measures show that galaxies in
the rest-frame UV are somewhat clumpier than rest-frame optical for both
galaxy populations. For star-forming galaxies at $z > 1.5$, \cite{bon11} and
\cite{law12} measured the internal color dispersion (ICD), and found that the
morphological differences between the rest-frame UV and optical are typically
small. However, the argument that the majority of ICDs for star-forming
galaxies are larger than those for passive galaxies \citep{bon11} is not
consistent with our finding of relatively large offset for pBzKs in
Figure~\ref{fig:zh}. Since most of our pBzKs ($\sim 80\%$) are massive ($M > 3
\times 10^{10}M_{\odot}$), one possibility is that high mass galaxies tend to
exhibit greater morphological differences with large ICD
\citep{law12}. Furthermore, pBzKs are typically brighter and rather compact at
rest-frame optical wavelengths, which results in higher $G$ and lower $M_{20}$
and $\Psi$ than in the rest-frame UV. On the other hand, \cite{szo11} found a
strong dependence of the morphology on wavelength in a visual study of 16
massive galaxies at $z\sim2$.

\section{Discussion}

In this section we briefly compare our measures of the morphologies of the mix
of galaxy populations in the redshift range $1.4 < z \le 2.5$ to the
predictions of theoretical models. In particular, we discuss the evidence that
the bimodal distribution of galaxies in the color-mass (or luminosity)
diagram, namely the ``red sequence'' and ``blue cloud'', has already started
to appear at $z\sim2$, and compare it with existing measures at lower
redshift.  The reliability of our non-parametric measures, $G$, $M_{20}$ and
$\Psi$ and their performance in quantifying the morphology of galaxies at
$z\sim2$, especially for less massive ones, are also discussed. Lastly, we
discuss the comparison between BzK and SED selected samples.

\subsection{ Comparison to the predictions of theoretical models}
Our analysis of the various morphological indicators, both parametric and
non-parametric ones, as well as a simple visual inspection
(e.g. Figure~\ref{fig:mont}), have shown that star-forming galaxies (sBzKs)
exhibit a broad variety of morphological structures, ranging from galaxies
with a predominant disk morphology and varying degree of bulge--to--disk ratio
to irregular (clumpy) structures to very compact and relatively regular
galaxies. Generally, the mix of star--forming galaxies at $z\sim 2$ looks
rather different from its counterpart in the local universe, showing a much
larger fraction of irregular and disturbed morphologies, especially among
massive and luminous galaxies, although bright galaxies that closely resemble
local spirals are also observed. We do observe luminous, clumpy galaxies whose
light profile is consistent with a disk (of course, we do not have dynamical
information on these galaxies) and whose overall morphology is in broad
qualitative agreement with the theoretical predictions of violent disk
instability (VDI, Dekel et al. 2009a) as seen in high--resolution hydrodynamic
cosmological simulations \citep{cev10}. These simulations show that galaxy
disks are built up by accretion of continuous, intense, cold streams of gas
that dissipate angular momentum in a thick, toomry--unstable disk, where
star--forming clumps form. Subsequently, clumps migrate toward the center and
edge, giving rise to bulges and pseudo--bulges. Observations of the
star--formation rate, stellar mass and age of the clumps, as well as their
average radial dependence relative to the center of the galaxies are also
broadly consistent with this scenario (e.g. Guo et al. 2011).

On the other hand, passive galaxies (pBzKs) are mostly spheroidal-like,
comparatively more regular and compact structures, a fact that has been
consistently observed in previous works \citep{dad05,fra08,van08, cas10,
  cas11}. It is important to keep in mind, however, that there is scant
spectroscopic information on the dynamical properties of these galaxies,
namely whether they are primarily supported by velocity dispersion or by
rotation. While the modicus of spectroscopic observations currently available
\citep{van11, ono11} is certainly consistent with the high--redshift passive
galaxies being spheoroids, a significant or even dominant contribution by
rotation cannot be ruled out given the limited angular resolution of the
existing spectra, and some have indeed proposed that a significant fraction or
maybe even most \citep{vdw11, bru12} of these galaxies are actually compact
rotating disks. From the theory point of view, there are three distict
scenarios for the formation of the compact spheroids, namely major mergers,
multiple minor mergers or the migration of clumps driven by violent disk
instabilities \citep{dek09b, gen11} to the disks center, building a massive
bulge. Cosmological hydrodynamical simulations indicate that these processes,
operating alone or in combination, can form compact spheroids. In most cases,
the inner parts of the compact spheroids formed by VDI are rotating, and the
outer parts are non-rotating, formed mostly by minor mergers. While the
observed morphological properties of galaxies certainly include cases that are
broadly consistent with these scenarios, it is clear that to make progress
comparisons between the dynamical properties of the galaxies and the
predictions of the simulations are necessary. These, however, require
spectroscopic observations with sensitivity and spatial resolution that are
not currently available.

In general very compact and massive galaxies are thought to be the result of a
highly dissipative process, either a major wet merger (e.g. Wuyts et~al. 2010)
or direct accretion of cold gas. Accretion of cold gas from the
inter--galactic medium \citep{bir03, ker05, dek06} can lead to the formation
of compact, massive galaxies, either via VDI in a compact disk \citep{dek09b}
or via direct accretion of the gas traveling directly to the galaxy center
rapidly and forming stars in--situ \citep{joh12}. Quenching of the star
formation subsequently takes place late when the supply of gas is halted. The
simulations suggest that cold accretion is naturally interrupted in dark
matter halos more massive $\approx 10^{12}$ M$_{\odot}$ once the shocked halo
gas become too hot to allow the cold flows to penetrate the halo before they
themselves get shock-heated \citep{dek06}, leading the formation of a massive
compact passive galaxy. Additional feedback mechanisms from star--formation
itself \citep{dia12} and AGN \citep{spr05} can also contribute to suppress the
accretion of cold gas.

As mentioned earilier, \cite{bru12} have studied the morphologies of massive
galaxies at $1<z<3$ in the CANDELS-UDF field using $S\acute{e}rsic$ and
bulge+disc models, finding that at $z>2$ massive galaxies are dominated by
disk-like structures and 25-40\% of quiescent galaxies have disk-dominated
morphologies.  Following their classification of disks, namely $n<2.5$ (even
though they also use bulge- to-total H-band flux ratio), we find that about
50\% of our passive galaxies (with $SSFR < 0.01~Gyr^{-1}$) have exponential
light profiles with $n < 2.5$, i.e consistent with exponential disks. These
roughly classified passive disks have mostly $G > 0.5$,~$M_{20} < 1.45,~\Psi <
1.0$, suggesting that they generally are not clumpy structures, and their
morphology is characterized by a central bright nucleus surrounded by low
surface brightness features. The presence of passive disks seems inconsistent
with models where galaxy morphology transforms from a disk structure into a
bulge followed by quenching of star formation as the galaxy evolves. The
existence of passive disks is, however, predicted by hydrodynamic simulations
\citep{ker05, dek08}, which show that these structures form when cold gas
inflows are halted, thus quenching star formation without the transformation
of morphology. For example, \citep{wil13} argue that the
morphological properties and volume density of massive, compact passive
galaxies at $z\sim 2$ and those of compact star--forming galaxies at $z>3$ are
generally consistent with such a scenario. 

\subsection{Bimodal color distribution at $z<2.5$}

In this study, to the extent that passive and star--forming galaxies can
effectively be identified by means of broad--band colors, e.g. either the BzK
selection criteria or via SED fitting, we have shown that passive (pBzK) and
star-forming (sBzK) galaxies occupy regions of the rest-frame U-V and stellar
mass diagram (e.g. Figure~\ref{fig:CM}) that are essentially the same as the
``red sequence'' and ``blue cloud'' observed in the local universe
(e.g. Blanton et al.\ 2003, Bell et al.\ 2004). Passive galaxies (pBzKs) are
intrinsically red and massive, whereas, star-forming galaxies (sBzKs) are
generally bluer and have lower mass than passive (pBzK) ones (with the
exception of about 7\% red massive sBzKs). The majority of the exceptions 
are massive dusty star-forming galaxies, and are largely redder than 
low mass star-forming galaxies (sBzKs) at $z\sim2$. Thus, they can 
contaminate the red sequence sample by a significant fraction, 
if selected based only on a single rest-frame (U-V) color \citep{bra09}, 
since most of the UV emission from high-redshift star formation is at least 
somewhat obscured by dust. 

An intriguing property of this color-mass diagram is the lack of passive
galaxies with mass $M<10^{10}$ M$_{\odot}$.  To a minor extent, this is the
result of incompleteness, since passive galaxies with lower mass, and hence
luminosity, become harder to detect. From the simulation using model galaxies
in the GOODS-S Deep field mosaics, we confirm that the early-type galaxies
with ``de Vaucouleurs" light profile ($S\acute{e}rsic$ Index = 4) are 90\%
complete with $H<26$. Clearly, however, such a small incompleteness
alone cannot explain the lack of low--mass passive galaxies at $z\sim 2$, and
in fact such low--mass galaxies are actually detected by the SED selection, as
shown in the right panel of Figure~\ref{fig:CM} which illustrates how massive
red galaxies are rarely star-forming, and more actively star-forming galaxies
are bluer and have lower masses. Furthermore, obscured star--forming galaxies
in the same redshift range and with similar rest--frame (U-V) colors are
detected in significantly larger number even at lower masses, as the right
panel of the figure shows. This fact strongly suggests that the quenching of
star-formation at this epoch is tightly correlated with the mass of the
galaxies, with the most massive ones being significantly more likely to cease
their star formation activity. At mass $M< 10^{10}$ M$_{\odot}$, galaxies
appear much less likely to stop forming stars, a fact that has been observed
by other groups. For example, Kauffmann et al.(2003) and Bell et al.(2007)
observed that at $z<1$ the stellar mass value of about
$3\times10^{10}M_{\odot}$ appears to be the transition mass point between
galaxies that belong to the blue cloud (younger stellar populations) and those
of the red sequence.

There is evidence that this transition mass between quenched and star-forming
galaxies has evolved significantly over cosmic time \citep{bun07} further
supporting the downsizing scenario whereby more massive galaxies appear to
quench first, and subsequently lower mass galaxies quench later. At high
redshift, quenching appears to depend on galaxy stellar mass, perhaps through
some internal process that is tied to the total mass of the galaxy of which
the stellar one is a good proxy in passive systems, whereas later,
environmental processes can contribute to galaxy quenching and can affect
lower-mass galaxies (e.g. Peng et al 2010, Peng et al 2012). This process
effectively builds up the lower-mass end of the red sequence over time by
quenching lower-mass star-forming galaxies later when environmental processes
become more influential. Our observed deficiency of lower-mass passive
galaxies at $z\sim2$ is consistent with this scenario and the implied
mechanisms by which galaxies quench their star-formation.

Results from other deep extragalactic surveys have provided constraints on the
buildup of stellar mass locked up in the red sequence, by studying how the
bimodality of galaxy properties has changed over cosmic time. For example,
using the COMBO-17 and DEEP2 surveys, \citep{bel04, fab07} have studied the
evolution in the rest-frame color bimodality of galaxy samples out to
$z\sim1$, finding evidence that the buildup of the red sequence must be
accounted for by a combination of merging of galaxies already on the red
sequence, as well as migration of star-forming galaxies that have
quenched. Recently, Brammer et al (2011) extended the study of rest-frame
color bimodality in galaxies to higher redshift, showing that star-forming and
passive galaxies are still robustly separated in color over the redshift range
$0.4 < z < 2.2$, and coming to the similar conclusion that the growth of the
red sequence must come from both merging and migration, particularly for
galaxies above the apparent quenching threshold, $M > 3\times10 M_{\odot}$. In
this context, a possibility is that that the morphological bimodality we have
observed in this study may imply that some degree of morphological
transformation must accompany the quenching of star-forming galaxies at $z <
1.4$, if they are to match the properties of the red sequence after
quenching. Regardless of the dominant mechanisms building the red sequence,
and weather or not the dominant mechanisms evolve with redshift, our result
suggests that the process was already underway at redshift 2. In other words,
at this epoch, the formation of the Hubble sequence is already
underway. Further detailed study of the evolution of morphological properties
of galaxies as a function of mass and star-formation properties will be
required to identify the specific mechanisms contributing to the growth of the
red sequence.

\subsection{ The reliability of non-parametric measures}

We mainly use the $G$, $M_{20}$ and $\Psi$ to study the morphologies of
galaxies at $z\sim2$. Those non-parametric measures are widely used to study
galaxy morphologies at high redshift, especially for large samples
\citep{law07, con08, law12, wan12}. We investigate the robustness of the $G$,
$M_{20}$ and $\Psi$ parameters in relation to the SNR in Appendix A and show
that any differences in the estimated parameters for the same galaxies
observed in the GOODS-S and UDF images, whose only difference is the vastly
different total exposure time, are relatively small. Most of the galaxies in
our samples ( above $90\%$ of BzK and $70\%$ of SED selected galaxies) have
reliable morphological measurements with $S/N_{pp} > 2.5$ at
$z\sim2$. Moreover, the good correspondence between those parameters and
visual inspection (in Figure~\ref{fig:mont}), as well as model-dependent
parameters indicates that our $G$, $M_{20}$ and $\Psi$ measures are not biased
by low signal-to-noise.  We note that $S\acute{e}rsic$ index alone is
generally not sufficient to quantify the morphology of low mass galaxies since
we find no correlation between $S\acute{e}rsic$ index and non-parametric
measures in the lower mass systems ($M < 10^{10}M_{\odot}$). Also
non-parametric measures more effectively characterize the morphology of
irregular galaxies (LPM04). Therefore, it is crucial to use non- parametric
diagnostics instead of, or at least in parallel with, the commonly used
$S\acute{e}rsic$ profile to study the morphology of lower mass galaxies and to
explore the origin of the Hubble sequence at $z\sim2$, and epoch when many
galaxies appear irregular.

\subsection{ The good performance of the BzK-selected sample}

Both the BzK and SED samples show very similar morphological distributions in
all the analysis done here. The comparison of the two samples confirms that
they are similarly representative of the mix of bright galaxies at
$z\sim2$. We additionally compare the distribution of morphological parameters
for the 136 spectroscopically confirmed BzK galaxies (8 pBzKs and 128 sBzKs,
specz sample) to the parent distribution (BzK sample). As expected, the
relative distributions of $G$, $M_{20}$ and $\Psi$ are similar to our parent
sample, and the median values of each parameter are almost the same, with the
exception of $G$ in the case of sBzKs. The average $G$ value of BzKs in the
specz sample is slightly higher, since over $70\%$ of the specz sample are
bright and massive ($M > 1\times 10^{10}M_{\odot}$) galaxies, which tend to
have higher $G$. Thus, in conclusion, the very similar results derived from
both samples proves the effectiveness of the BzK selection criteria in
sampling the full diversity of the mix of massive galaxies at $z\sim 2$, at
least as far as the morphological properties of relatively massive galaxies
are considered. The BzK selection will be particularly effective, for example,
in the other three CANDELS fields where the broad--band photometry is neither
as deep or as broad in wavlength as in the two GOODS fields.

\section{Summary}

In this paper we have explored general trends between galaxy morphology and
broad--band spectral types at $z\sim2$ using the {\it HST}/WFC3 H--band images
taken in the GOODS--South field as part of the CANDELS survey, in combination
with the existing GOODS ACS z-band images, as well as sensitivity--matched
images at other wavelngths that are part of the GOODS data products
\citep{gia04}.  Combining the deep and high-resolution NIR data to optical
data, we are able to study the dependence of morphologies on wavelengths and
expand the scope of previous studies of galaxy morphologies at the same
redshift \citep{kri09, cam11, szo11, wan12} with significantly larger sample
size and lower mass limit ($>10^{9}M_{\odot}$). The galaxies of our primary
sample are selected in the redshift range $1.4<z<2.5$ to cover a broad range
of spectral types (star--formation properties and dust obscuration) using SED
fitting to spectral population synthesis models (SED sample). For comparative
reasons we also selected galaxies using the BzK creterion which culled
galaxies in the same redshift range and with essentailly identical spectral
properties, modulo a large contamination from interlopers and AGN. Analyses of
the two samples show consistent results suggesting that the BzK and SED
selection criteria are equivalent in sampling the mix of spectral types at
$z\sim 2$. We investigate the rest-frame optical morphologies using five
non-parametric approaches, mainly $G$, $M_{20}$ and $\Psi$ in addition to $C$
and $A$, and two model-dependent parameters obtained by fitting
$S\acute{e}rsic$ profiles, namely $n$ and $R_{e}$. The major findings of this
study are presented below.

\begin{enumerate}

\item In the rest-frame (U-V) color and mass diagram, our sample clearly
  separates into red massive passive galaxies with low SSFR and blue
  star-forming ones with less massive, high SSFR occupying the same regions in
  the color-mass diagram as the galaxies observed in the local universe.

\item We find that galaxies with different spectral types are distinctly
  classified morphologically as two populations, especially for massive
  systems ($>10^{10}M_{\odot}$) : 1) star-forming galaxies are heterogeneous,
  with mixed features including bulges, disks, and irregular (or clumpy)
  structures, with relatively low $G$, $n$ and high $M_{20}$, $\Psi$; 2)
  passive galaxies are spheroidal-like compact structures with higher $G$, $n$
  and lower $M_{20}$ and $\Psi$. Generally, the sizes of star-forming galaxies
  are larger than passive ones, even in massive systems, but some have very
  compact morphologies, with $R_{e} <1kpc$. We confirm using a variety of
  measures that star formation activity is correlated with morphology at
  $z\sim2$, with the passive galaxies looking similar to local passive ones
  although smaller, while star-forming galaxies show considerably more
  mophological diversity than massive star-forming galaxies on the Hubble
  sequence today.

\item We show that the morphological analysis only using the $S\acute{e}rsic$
  index is not sufficient to charaterize differences in morphologies
  especially for lower mass galaxies ($M<10^{10}M_{\odot}$). Therefore we
  conclude that it is important to use non-parametric measures to investigate
  the morphologies of high redshift galaxies in a broad range of stellar
  masses. In this study, the combination of large samples with a suite of
  morphological diagnaostics, both parametric and non--parametric ones, as
  well as visual inspections, gives us a significantly improved description of
  the state of galaxy morphologies at $z\sim2$ and its correlations with the
  spectral type, i.e. mostly the star--formation activity, expanding the
  significance and the scope of previous studies which were based on much
  smaller samples and only massive galaxies at the same epoch
  \citep{kri09,cam11, szo11, wan12}.

\item Generally, $z\sim2$ galaxies show a similar trend in morphologies with
  those measured from the redshifted images of local galaxies, even though
  many of the star-forming galaxies have $M_{20}$ values higher than seen for
  galaxies in the local sample. The passive galaxies at $z\sim2$ have $G$ and
  $M_{20}$ values that are much more similar to those observed for the
  present--day E/S0 galaxies.

\item Comparison of visual images between the rest-frame optical and near UV
  show that the morphological k-correction is generally weak, however, the
  comparison with non-parametric measures indicates that galaxies observed in
  the rest-frame UV are slightly clumpier, with lower $G$ and higher $M_{20}$
  and $\Psi$, than rest-frame optical.
\end{enumerate}

Taken all together, our results show that the correlations between morphology
as traced by a suite of common diagnostics, and broad--band UV/Optical
spectral types of the mix of relatively massive galaxies (i.e. $M>10^{9}$
M$_{\odot}$) at $z\sim 2$ are quantitatively and qualitatively similar to
those observed for their counterparts in the local universe. We interpret
these results as evidence that the backbone of the Hubble Sequence observed
today was already in place at $z\sim 2$.

\acknowledgements
This work is based on observations taken by the CANDELS Multi-Cycle Treasury 
Program with the NASA/ESA HST, which is operated by the Association of 
Universities for Research in Astronomy, Inc., under NASA contract NAS5-26555.

\appendix

\begin{figure}
\epsscale{0.8}
\plotone{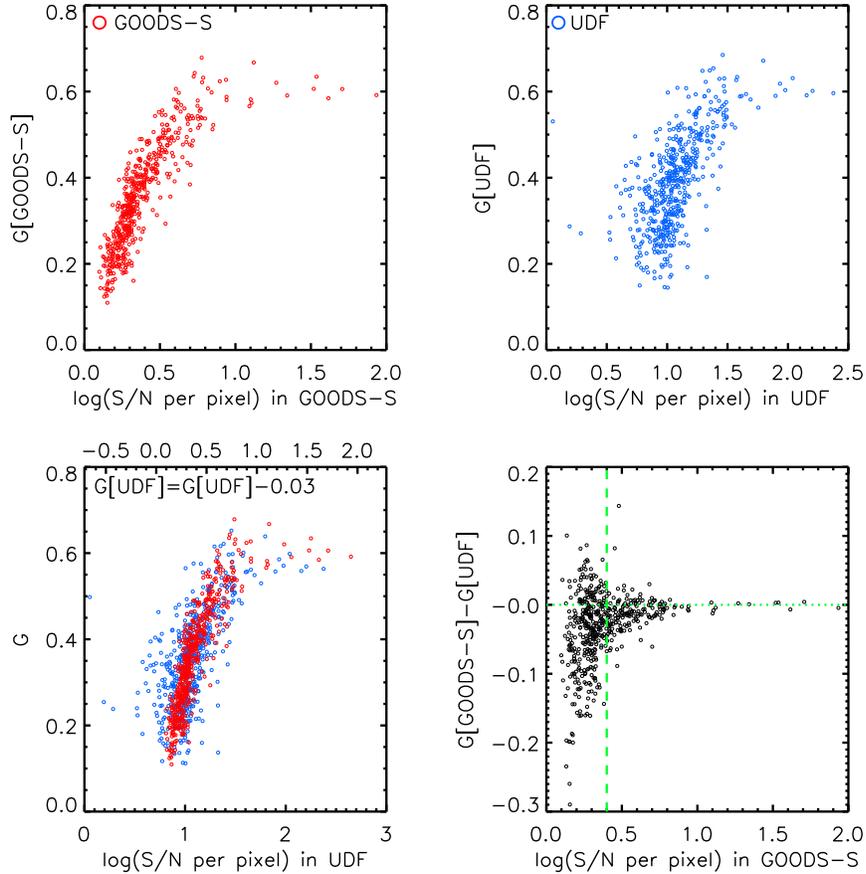}
\caption{Gini vs. logarithmic signal-to-noise per pixel for the same galaxies imaged with 
differing image depths in UDF (blue) and GOODS-S (red). In the bottom left panel, we have shifted 
G values in UDF with a median of differences, $(G[GOODS]-G[UDF])=-0.03$. The bottom right 
panel is the plot of log(S/N per pixel) vs. the difference in G between GOODS and UDF. 
The green dashed line at log(S/N per pixel)=2.5 indicates the signal-to-noise limit for 
robust measures discussed in the text.}
\label{fig:gini_sn}
\end{figure}

\section{Robustness of Parameter Estimations}

The reliability of model-independent parameters has been tested and discussed 
by many previous studies \citep{lot06,lis08,and11, law12}. These parameters are 
robust for large bright sources (with high signal-to-noise ratio) in general, but it 
can be unreliable for small faint sources. In particular, the $G$ value is very sensitive 
to the signal-to-noise ratio (LPM04 and Lisker et al. 2008). Our galaxy sample 
goes deeper ( $H < 26$), and to lower stellar mass limits ($\sim 10^{9}M_{\odot}$)
 than other works that study morphologies at $z\sim2$ 
(e.g. Kriek et al. 2009, Szomoru et al. 2011, Wang et al. 2012). Therefore, 
it is crucial to understand how noise and other limitations affect parameters 
we use here. We test the dependence of the parameters, $G,~M_{20}$ and $\Psi$, 
on the signal-to-noise ratio, i.e., the depth of an image. HST/UDF WFC3 imaging 
\begin{figure}
\centering
\epsscale{0.8}
\plotone{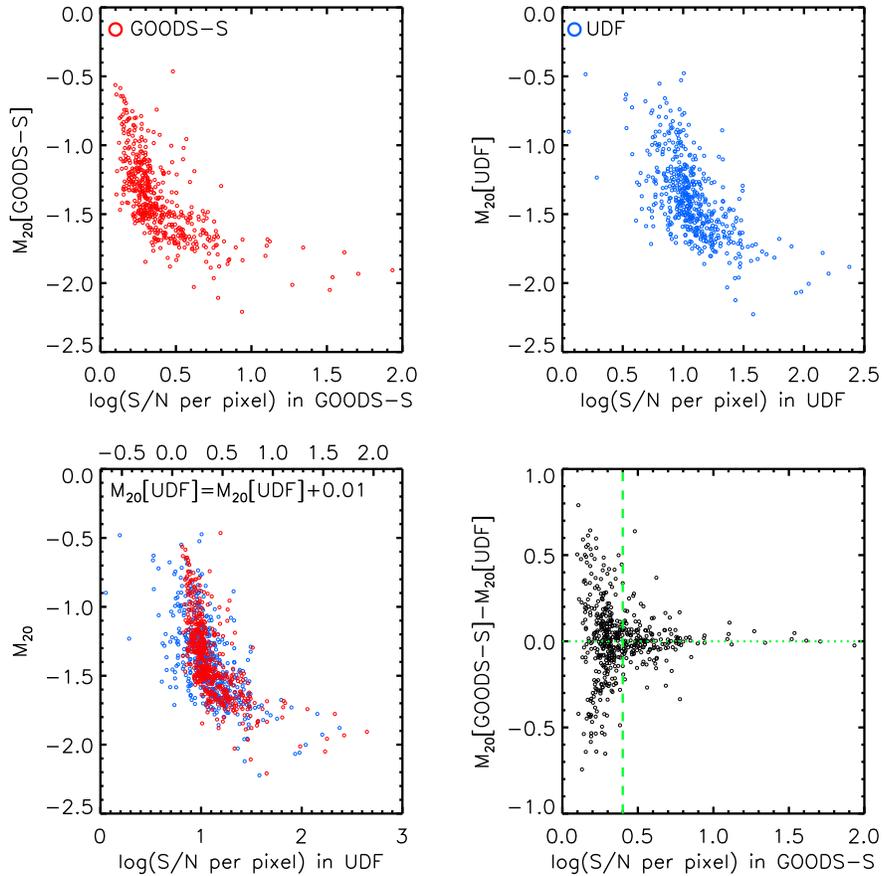}
\caption{$M_{20}$ vs. logarithmic signal-to-noise per pixel for the same galaxies 
imaged with differing image depths in UDF (blue) and GOODS-S (red). In the bottom 
left panel, we have shifted $M_{20}$ values in UDF with a median of differences, 
$(M_{20}[GOODS]-M_{20}[UDF])=0.01$. The bottom right panel is the plot 
of log(S/N per pixel) vs. the difference in $M_{20}$ between GOODS and UDF. 
The green dashed line at log(S/N per pixel)=2.5 indicates the signal-to-noise 
limit for robust measures discussed in the text.}
\label{fig:m20_sn}
\end{figure}

\begin{figure}
\epsscale{0.8}
\plotone{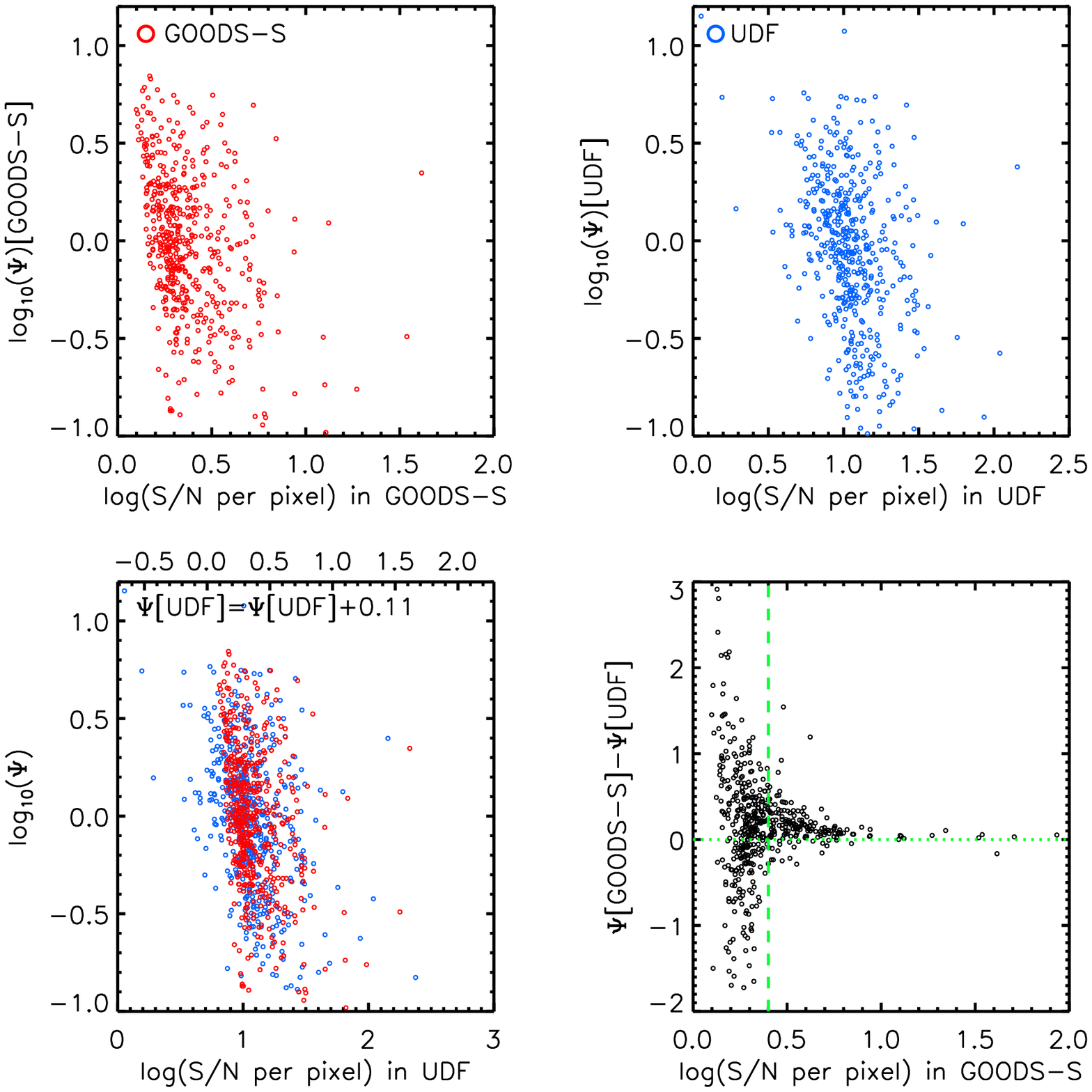}
\caption{$\Psi$ vs. logarithmic signal-to-noise per pixel for the same galaxies imaged 
with differing image depths in UDF (blue) and GOODS-S (red). In the bottom 
left panel, we have shifted $\Psi$ values in UDF with a median of differences, 
$(\Psi[GOODS]-\Psi[UDF])=0.11$. The bottom right panel is the plot of log(S/N per pixel) vs. 
the difference in $\Psi$ between GOODS and UDF. The green dashed line at 
log(S/N per pixel)=2.5 indicates the signal-to-noise limit for 
robust measures discussed in the text.}
\label{fig:psi_sn}
\end{figure}
data, which differ only in exposure time from GOODS-S data, is an ideal 
comparison dataset for investigating the effect of the different 
image depth on these parameters. Previously, Lisker et al. 2008 tested the dependence of 
signal-to-noise on $G,~M_{20}$ in the UDF and GOODS-S using $i$-band images. 
To investigate how this affects our measurements with the NIR WFC3 images, 
we use the UDF and shallower GOODS-S F160W (H-band) data (both with 
60mas pixel scales) and select 959 sources in the UDF image using GOODS-S 
source catalog. Then, we estimate $a_{p}$ from the GOODS-S H-band image, and 
use it to measure the parameters for both the UDF and GOODS-S images. The final 
sample is 520 objects which have stellarity $\le 0.8$, after excluding sources for 
which the number of pixels within $a_{p}$ is less than 28 pixels 
(same exclusion with Lisker 2008).
Signal-to-noise ratio per pixel ($S/N_{pp}$) is defined as
\begin{equation}
S/N_{pp}=\frac{1}{N_{pix}}\sum^{N_{pix}}_{i}\frac{f_{i}t_{exp}}{\sqrt{rms_{i}t_{exp}+f_{i}t_{exp}}}
\end{equation}
where, $N_{pix}$ indicates the total number of pixels within the $a_{p}$, $t_{exp}$ 
is the total image exposure time. $f_{i}$ and $rms_{i}$ are the pixel values in the 
drizzled image and rms image, respectively. The strong dependence of 
$G,~M_{20},~\Psi$ on the $S/N_{pp}$ are shown in figures~\ref{fig:gini_sn}, 
\ref{fig:m20_sn}, and \ref{fig:psi_sn}, respectively. 
The two top panels in figure~\ref{fig:gini_sn} present how values of $G$ vary 
with $S/N_{pp}$ in the GOODS-S (red circles) and UDF (blue circles). 
Obviously, the UDF has a higher $S/N_{pp}$ (difference of depth is about 
1.8 mag), but the overall trend of $G$ looks similar to that for GOODS-S. 
Both $G$ values overlap almost exactly when we move the Gini of the UDF 
to the relatively small median difference (-0.03) shown in the bottom left 
panel. However, the $G$ of galaxies with different exposure times is dependent 
on the $S/N_{pp}$ as shown in the bottom right panel. The difference of 
in $G$ between the UDF and GOODS-S decrease as $S/N_{pp}$ increases. 
$M_{20}$ and $\Psi$ also follows the same trend and distribution as $G$ as 
shown in figure~\ref{fig:m20_sn} and \ref{fig:psi_sn}, but with small median 
differences of about 0.01 and 0.11, respectively. Consequently, the 
model-independent parameters depend on signal-to-noise ratio, but the 
difference between two fields with different exposure times are less significant
 than \cite{lis08}. LPM04 find that $G$ and $M_{20}$ are reliable for galaxy images 
with $S/N_{pp} \gtrsim 2.5$ (vertical dashed line on the bottom right panel 
in each figure). Given this threshold in reliability, we investigate how this may 
affect our results by measuring the $S/N_{pp}$ of each galaxy. We find over 
90\% of BzK sample and 70\% of SED sample having $S/N_{pp} \ge 2.5$, 
which excludes some faint star-forming galaxies and passive galaxies with low $G$ 
from our sample. Repeating our analysis with these faint objects removed 
doesn't change the results. Therefore, we confirm that the results from our 
measurements did not suffer from this $S/N$ effect.

\clearpage

\end{document}